\documentclass{llncs}
\usepackage{mathptmx}       
\usepackage{helvet}         
\usepackage{courier}        
\usepackage{makeidx}         
\usepackage{graphicx}        
\usepackage{multicol}        
\usepackage[bottom]{footmisc}
\usepackage{subfigure}
\usepackage{amsfonts,dsfont}
\usepackage[cmex10]{amsmath}
\usepackage{cite}
\usepackage{epsfig,amssymb,amsmath, multirow, url, tcolorbox,booktabs, enumitem}
\usepackage{epsfig, amssymb, multirow, url, tcolorbox,booktabs}
\usepackage[figuresright]{rotating}
\usepackage{colortbl}

\usepackage{algorithm, algorithmicx}
\usepackage[colorlinks=true,linkcolor=cyan,citecolor=blue]{hyperref}
\usepackage[utf8]{inputenc}
\usepackage{bbm}
\usepackage{authblk}
\usepackage{algpseudocode}
\algrenewcommand\algorithmicrequire{\textbf{Input:}}
\algrenewcommand\algorithmicensure{\textbf{Output:}}

\usepackage{amsmath}

\newtheorem{@assumption}{\sc Assumption}[section]

\newcommand{\wha}{\widehat{\alpha}}
\newcommand{\whb}{\widehat{\beta}}
\newcommand{\whs}{\widehat{\sigma}^2}
\newcommand{\whr}{\widehat{\rho}}
\newcommand{\whP}{\widehat{\Psi}}
\newcommand{\che}{\check{\epsilon}}
\newcommand{\whe}{\widehat{\epsilon}}
\newcommand{\what}{\widehat{\alpha}^{*}}
\newcommand{\whbt}{\widehat{\beta}^{*}}
\newcommand{\whst}{\widehat{\sigma}^{*}}
\newcommand{\whrt}{\widehat{\rho}^{*}}
\newcommand{\whPt}{\widehat{\Psi}^{*}}

\newcommand{\whatc}{\widehat{\alpha}^{(\text{cand})}}
\newcommand{\whbtc}{\widehat{\beta}^{(\text{cand})}}
\newcommand{\whstc}{\widehat{\sigma}^{2,(\text{cand})}}
\newcommand{\whrtc}{\widehat{\rho}^{(\text{cand})}}
\newcommand{\whPtc}{\widehat{\Psi}^{(\text{cand})}}

\newcommand{\wtp}{\widetilde{\Phi}}

\newcommand{\fc}{\mathfrak{c}}

\newcommand{\nett}{\texttt{nett}}
\newcommand{\RR}{\texttt{R}}

\newcommand {\argmin}[1]{\underset{#1}{\rm{argmin}}} 
\newcommand {\argmax}[1]{\underset{#1}{\rm{argmax}}}

\newcommand{\harpoon}{\overset{\rightharpoonup}}

\usepackage{todonotes}

\definecolor{Gray}{gray}{0.85}

\title{ACRONYM: Augmented
degree corrected, Community Reticulated Organized Network Yielding Model}

\titlerunning{ACRONYM}
\author{Benjamin Leinwand\inst{1} \and Vince Lyzinski\inst{2}
	}
 \authorrunning{Benjamin Leinwand and Vince Lyzinski}

\institute{Stevens Institute of Technology\\
		\and
	University of Maryland, College Park}

\begin{document}

\maketitle
\begin{abstract}
{Modeling networks can serve as a means of summarizing high-dimensional complex systems. Adapting an approach devised for dense, weighted networks, we propose a new method for generating and estimating unweighted networks. This approach can describe a broader class of potential networks than existing models, including those where nodes in different subnetworks connect to one another via various attachment mechanisms, inducing flexible and varied community structures. While unweighted edges provide less resolution than continuous weights, restricting to the binary case permits the use of likelihood-based estimation techniques, which can improve estimation of nodal features. The extra flexibility may contribute a different understanding of network generating structures, particularly for networks with heterogeneous densities in different regions.}

\end{abstract}

\section{Introduction}
When observing a network $G =\{V, E\}$, represented as a symmetric adjacency matrix $A$, without having access to any additional information such as nodal attributes, we seek to infer the mechanism that generated the network. 
Many large scale real-world networks have been observed to have heavy tailed degree distributions, where certain nodes have far more incident edges than expected given the average node degree. 
Though more recent work \cite{broido2019scale} has cast doubt on the ubiquity of ``scale-free" networks, it is imperative to account for the variation of node degrees when analyzing a network \cite{DCBM}. 
Another fundamental property of many networks is the presence of communities, in that all nodes in the network do not behave homogeneously, but instead, the network can be partitioned into subsets of nodes which seem to exhibit common behaviors \cite{fortunato2016community,girvan2002community}.

Many methods and algorithms exist for detecting the community structure and generative mechanism of a network; see, for example, the survey on the myriad applications/estimations of stochastic blockmodels (SBMs) \cite{abbe2018community}.
For our purposes, the problem of inferring the network generative process can be subdivided into estimating three sets of quantities; note that these parameters are similar in spirit to those in the degree-corrected SBM \cite{DCBM} and popularity adjusted SBM \cite{PABM}. 
First is the community membership of each node, if such communities exist.
Second is the ``sociability" of the node, a measure of degree heterogeneity that reflects each node's social proclivity. 
This ``sociability" need not be constant over the whole network, as a node may be more or less prone to connecting with members of various communities.
Finally, we wish to recover the functions modeling edge formation probabilities based on the incident nodes' sociabilities and community memberships. 
In the case where there are few communities relative to the number of nodes in the network ($K\ll N$, where here $K$ represents the number of communities and $N$ the number of nodes), and when the edge formation probabilities are based solely on node sociabilities and community memberships, then the problem of estimating $N(N-1)/2$ edges can be reduced to estimating $O(NK)$ different parameters; such a reduction in model complexity is key for estimation, and is a common theme throughout the network estimation literature \cite{RDPG, durante2017nonparametric,rubin2022statistical}. 
In the case where it can be further assumed that each node's sociability is in fact constant everywhere in the network, as in the degree-corrected SBM, the same edges can be estimated with only $O(N+K)$ parameters.
Reducing the problem also provides insight into the generative process for the network structure, in that while each edge is often assumed to be \textit{conditionally} independent (given the latent parameters of the model) of other edges, the edges are not independent random variables.
Rather, latent information---here, in the form of these three parameters---is embedded in the whole network, and can be leveraged to better understand the probability that any potential edge will be present.

There are several existing models for generating unweighted networks, including some which allow for community structures as well as differing nodal sociabilities amongst various communities. 
Prominent examples include the degree corrected stochastic blockmodel (DCBM) \cite{DCBM} and the popularity adjusted blockmodel (PABM) \cite{PABM, noroozi2021estimation, noroozi2021sparse}, both of which extend the classical stochastic blockmodel (SBM) \cite{holland1983}.
While myriad other SBM variants and extensions have been posited in the literature (see, for example, \cite{airoldi08,peixoto2014hierarchical,li2022hierarchical}), we focus here on the DCBM and PABM, as they bear the closest relation to our posited ACRONYM model introduced in Section \ref{s-unweightedmodel} (ACRONYM for ``Augmented
degree corrected, Community Reticulated Organized Network Yielding Model").
In the classical SBM, the node set is partitioned into communities, and edges are conditionally independent given the community memberships of the nodes.  Moreover, the probability of two nodes forming an edge is solely a function of their respective community memberships.
The degree corrected stochastic blockmodel (DCBM) \cite{DCBM} incorporates the community structure of the SBM and allows for certain nodes to have uniformly greater connection probabilities than other nodes.
In the DCBM, each node is assigned a degree-correction parameter, and the probability of two nodes forming an edge is a function of their respective community memberships and their degree correction parameters. 
However, in the DCBM, more ``social" nodes are globally more social than less social nodes. 
The popularity adjusted blockmodel (PABM) relaxes this assumption, allowing one node to be relatively more or less popular than another depending on the community membership of the node to which they wish to connect. However, PABMs still have a fairly restrictive form in terms of permissible functions for defining edge probabilities. 

\begin{figure*}[t!]
\includegraphics[width=\linewidth]{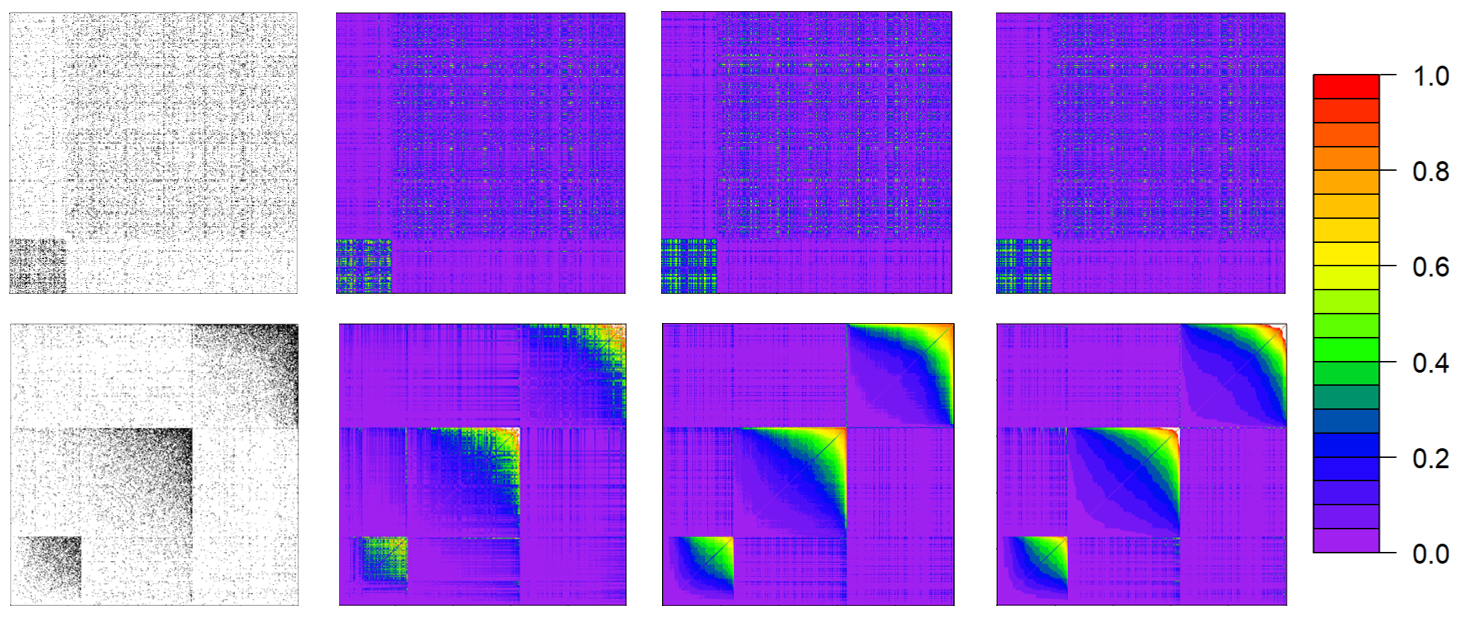}
\caption{Top row from left to right: the binarized, undirected network of Twitter interactions between 475 members of the US Congress between February 9, 2022 and June 9, 2022, where the first 92 nodes (starting from the bottom left) are senators, and the rest are congresspeople. The estimate of the probability of connecting using the Degree Corrected Block Model estimated using the $\nett$ $\RR$ package. The predictions of the probability of connecting from our proposed method using 10-fold cross-validation. The estimate of the probability of connecting using the Popularity Adjusted Block Model but the community detection approach from our model. The bottom row shows the same information as the top row, but the with nodes reordered first by the communities recovered using our community detection approach on the entire network, then by within community degree.}
\label{f:congress-estimates}
\end{figure*}

Even for those approaches where model fitting is relatively straightforward, capturing the structure of the network in the generative model can be nuanced and difficult. 
Consider, as a motivating example, the network in Figure \ref{f:congress-estimates}.
This network, originally described in \cite{fink2023centrality}, represents the Twitter interactions between the 475 members of the 117th U.S. Congress who issued at least 100 tweets between February 9, 2022 and June 9, 2022 \cite{fink2023twitter}. 
Though the original data is a directed, weighted network, for our purposes, we convert this to a network of unweighted, undirected edges. 
Starting from the bottom left corner, the first 92 positions in both dimensions of the plot represent senators, while the rest are congresspeople (i.e., members of the U.S. House of Representatives). 
On cursory examination of the adjacency matrix of the graph, one observes that certain members of congress are more connected than others, and that there are certain group dynamics at play. 
On the former point, even though edges in this network only take on values in $\{0, 1\}$, it is reasonable to infer that, ceteris paribus, the probability of observing an edge incident to a more connected node is greater than the probability of observing an edge incident to a less connected node. 
On the latter point, this particular view of the network (which follows the ordering of the data from \cite{fink2023centrality}) demonstrates that, per capita, there is more interaction between senators than between senators and congresspeople. However, this view obscures a different potential divide, namely that between the Democratic and Republican congresspeople (note that the Democratic and Republican parties are the two most common political parties represented in this Congress). 
One could imagine a number of other such salient hypothesized partitions (e.g. region, demographics; indeed there are often multiple ``true'' clusterings in a given network \cite{priebe2019two}) which might predict connectivity patterns in this network, though a priori knowledge of these nodal features goes beyond what is strictly observed in the network adjacency representation itself. 
Though examining the impact of these different features may be of interest to a political scientist, this paper assumes that only the network is observed, without observing other identifying node features. 
Given that, we seek to learn the community structure of the network, as well as the patterns of behavior exhibited by these different communities.

In the conditionally independent edge paradigm (where we assume that edges appear in the network conditionally independent given latent network parameters), we can view the modeling process as serving the ultimate goal of finding good estimates of the edge probabilities (often as functions of the network model parameters). 
In this motivating example network, an SBM is inappropriate because it would ignore any node to node variations in sociability beyond the community level, failing to capture (for example) if a member of Congress is a particularly prolific or meager tweeter. 
While the DCBM and PABM can capture this degree heterogeneity better, here the DCBM or PABM can actually predict that the probability of an edge between a particular pair of nodes exceeds 1. 
This is visible, especially in the case of the PABM,  in the in the bottom row of Figure \ref{f:congress-estimates}.
In that row, the 2$^{\text{nd}}$ and 4$^{\text{th}}$ panel represent the edge probability matrices estimated (using the \texttt{nett} package in \texttt{R}) by the DCBM and PABM respectively. 
In the PABM panel, note the trend in the top right corner of the within community edge probability estimates in the reordered networks, where values in white are greater than 1.     
In the DCBM panel, this is less prevalent (though still present), though the estimated edge probability matrices are less able to capture the transition from low-degree to high-degree nodes in the community.  
In the 3$^{\text{rd}}$ panel, where we fit the ACRONYM model, we see the transition from low-degree to high-degree nodes is captured well without the problematic tendency of the PABM to estimate edge probabilities with values greater than (or equal to) $1$.

\begin{figure}[h!]
\begin{center}
\includegraphics[width=140mm]{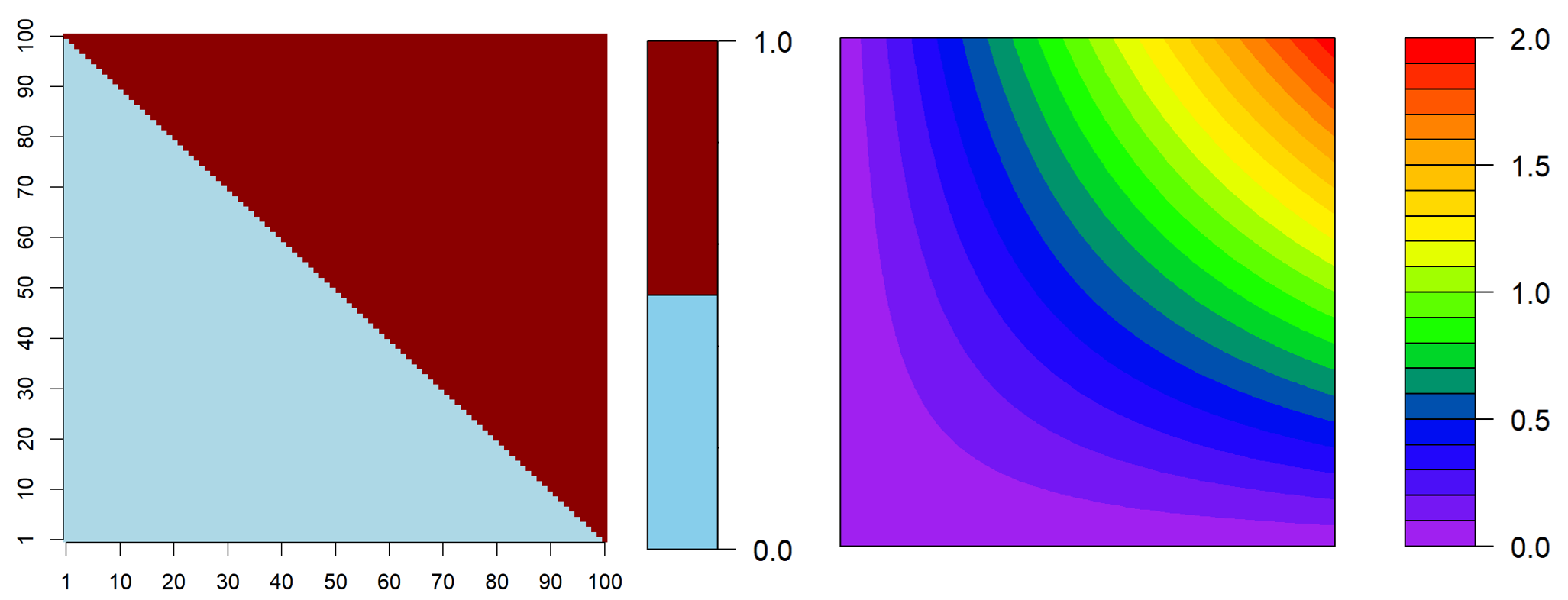}
\caption{\label{f:threshold}
Left: a toy subnetwork consisting of 2 communities of 100 nodes each, with a threshold. Light blue represents non-edges, dark red are observed edges. Right: DCBM estimates of the toy subnetwork assuming the nodes are correctly clustered.}
\end{center}
\end{figure}

To gain better insight here, we consider the simple synthetic example constructed in Figure \ref{f:threshold}. 
This is a subnetwork generated as follows: there are 2 distinct communities, each containing 100 nodes numbered 1 through 100; this network represents the edges between the two communities. If two nodes in different communities have numbers whose sum exceeds 100.5, an edge is placed between these nodes, else there is no edge.
Notably, this subnetwork does not follow the DCBM or PABM model. 
We further assume that nodes are clustered into the correct communities (without further subdividing of these communities), and that each node's degree proportion in this subnetwork perfectly reflects its degree proportion in the broader network. 
Under these assumptions, in the DCBM or PABM, each node $u$ would be assigned a parameter $\theta_u$ and there is a block parameter given by $\beta$. There are a total of 5050 edges in the subnetwork, where node 100 in each community is incident to 100 of the 5050 edges, node 99 in each community is incident to 99 of these 5050 edges, etc. Therefore, according to the DCBM or PABM, the estimated edge probability between the nodes labeled 100 in each community is given by (where $\hat\theta$ and $\hat\theta'$ are the estimated degree correction parameters for the two nodes and $\hat \beta$ the estimated block parameter)
$$\hat\theta\hat\beta\hat\theta'=\frac{100\cdot 100}{5050}\times\frac{5050}{100\cdot 100}\times\frac{100\cdot 100}{5050} = 1.980198.$$ Since this is modeling a network with unweighted edges, one would like the estimated edge probability for the single edge occurrence to be in $[0,1]$, which fails under the calculated DCBM estimate.
Furthermore, as the estimates in the right plot of Figure \ref{f:threshold}  show, the DCBM model fit to the block in this toy example would produce a large number of these impossible estimates, and the model assumes a convex contour structure that doesn't appear in the toy example. The DCBM was initially developed in \cite{DCBM} under the assumption of Poisson edges, a context in which this estimate would be acceptable, but it has become popular for modeling unweighted networks, a context where this estimate is often not sensible \cite{jin2015fast, zhao2012consistency, Hwang2023, lei2015consistency}. In these cases, valid probabilities are usually assumed along with a maximum sociability parameter for identifiability, but if the mechanisms generating real networks do not follow the structure of a DCBM, these models can yield incoherent probability estimates.

\begin{remark}
One might be tempted to attribute the result highlighted in Figure \ref{f:threshold} to the high density of the toy example.
However, the issue can arise when there are denser regions in a relatively sparser network. For example, the congressional Twitter network is only about 9\% dense. Rather than global density, it is the combination of density and wide degree distributions within subnetworks induced by pairs of communities that can produce this kind of flawed estimate. In those settings, DCBMs and PABMs struggle to achieve global network estimates which are locally admissible everywhere. 
\end{remark}

In this paper, we introduce a different approach (the ACRONYM model) that can model a broad set of networks with structures like those in the congressional network, without succumbing to the degeneracy seen in DCBMs. 
It can be thought of as an adaptation of \cite{LeinwandPipiras} to unweighted networks. 
The work in \cite{LeinwandPipiras} describes a model for generating dense weighted networks of the kind seen in structural brain networks. It also discusses methods of community detection, as well as estimation for those networks. The key insight of that work is that many established network science methods are built for a sparse regime, where most nodes are not connected, and so they cannot be seamlessly adapted to the case where all nodes are connected in weighted networks. The paper instead focuses on defining ``communities" based on an order of preference over nodes and potentially different patterns of edge weights, and it describes how to generate these different kinds of flexible connectivity patterns. 
While these results are perhaps more intuitively motivated in weighted networks with arbitrary edge weight distributions, there are aspects of that model that can still be brought to bear on unweighted networks, though they may manifest in different ways. For networks with unweighted edges, the chief contribution is finding a means of constructing a  variety of edge connectivity patterns without violating rules of probability. As in the weighted case, though, the notion that the connectivity ``contours" may take on different kinds of patterns in different communities (dubbed ``augmented degree correction"), can still apply to networks with unweighted edges. The  adjustment is that now these flexible contours explicitly appear only in the probability matrix underpinning the observed network. Consequently, the usual definition of a community as ``a subset of nodes with greater internal propensity to connect" can also be replaced by the alternative definition of a community being composed of a subset of nodes that share an order of preferences over other nodes. This new definition holds even in unweighted networks, even though preferential orderings are harder to assess given only two possible outcomes for any potential edge in the network. 
   
\section{$H$-functions}
\label{s-DenseModel}
The work in \cite{LeinwandPipiras} approaches the problem of modeling dense, weighted networks by studying the patterns/orderings of edge weights which are allowed to depend on particular nodes and communities. 
By focusing on the ordering of edge weights rather than the values of the edge weights themselves, the methodology is robust and can generate networks with flexible underlying sociability patterns (SC) and edge weight distributions that vary based on the communities being connected, while incorporating error terms whose magnitudes also depend on the communities. 
This enables heterogeneous modeling of dense systems where nodes in different ``modules" connect to one another via various attachment mechanisms. The work in \cite{leinwandrec} further adapts this approach for bipartite weighted networks for use in recommender systems. 
Herein, we modify this approach to model unweighted networks. 
Before we provide our novel ACRONYM model, it is useful to review some of the features of the overall approach. 

A core idea of that model is so-called $H$-functions, which can be understood as functions (similar in spirit to graphons \cite{lovasz2012large}) which take in two Unif$(0, 1)$ random variables, and output one Unif$(0, 1)$ random variable. 
A particular focus is $H$-functions with \emph{positive association}, or those $H$-functions that are monotonically increasing in both arguments. By contrast, $H$-functions with \emph{negative association} are decreasing in both arguments, and those with \emph{Simpson association} are increasing in one argument and decreasing in the other. 
Restricting our purview to a particular subnetwork between communities $c_i$ and $c_j$ (with $i=j$ allowed), the model posits that certain nodes in each community are more popular with nodes in the other community. Each node's local popularity can be reflected in their $\Psi$ value, their ``sociability" parameter. 
The SC in the subnetwork can be expressed as $[H(\Psi_u, \Psi_v)]_{u\in c_i, v\in c_j}$, which is composed of an $H$-function with positive association that takes the $\Psi$ values of the nodes being connected as inputs. 
When translating this SC to the ``normal space" in order to easily incorporate error, and then translating back to the ``edge weight space," the edges are generated as follows:    
\begin{equation}\label{e:HNSM}
 \Phi^{-1}(G(W_{uv})) = \frac{1}{\sqrt{1+\sigma^2}}  \Phi^{-1}(H(\Psi_u, \Psi_v))+ \frac{\sigma}{\sqrt{1+\sigma^2}} \epsilon_{uv},
\end{equation}
where $G$ is the CDF of the edge weights in that subnetwork, $\Phi^{-1}$ is the inverse CDF of the $\mathcal{N}(0, 1)$ distribution, $\epsilon_{uv}$ are i.i.d.\  $\mathcal{N}(0, 1)$, and $\sigma \ge 0$. 

\begin{figure}[t!]
\includegraphics[width=\linewidth]{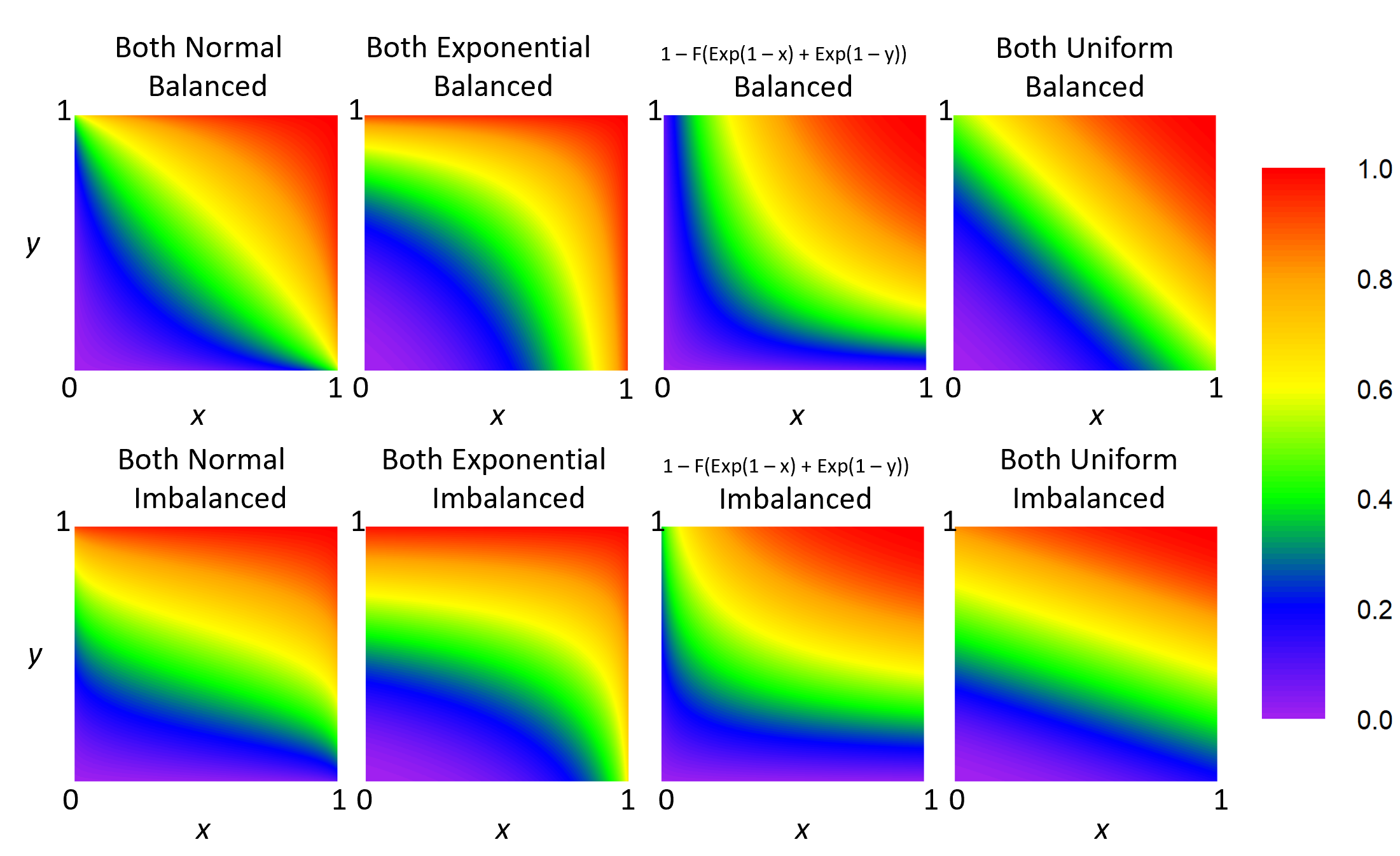}
\caption{Plots of examples of $H$-functions with positive association. From left to right on top row, $H$-functions with $\rho=1$ displaying ``normal," ``concave," ``convex," and ``linear" contours.  On the bottom row, these same $H$-function family, but this time with the value of $y$ being more influential on the final outcome than the value of $x$. In either row, any value $H_3(x, y)$ in the third column is equal to $1-H_2(1-x, 1-y)$, where $H_2$ represents the relevant $H$-function in the second column. The "Exp" in the third column represents the inverse CDF of an Exponential distribution.}
\label{f:different-h-funcs}
\end{figure}

While there is additional randomness injected due to the $\epsilon$ values, the expected ordering of the edge weights follows the ordering of $H(\Psi_u, \Psi_v)$, forming distinct degree correction patterns. In other words, $H$ serves as a map from orderings of nodal local popularities to orderings of edge weights. In cases where both $\Psi_u$ and $\Psi_v$ are large (or small), the edge weight would also be expected to be large (small). However, in a case where $0 < \Psi_a < \Psi_b < \Psi_c < \Psi_d < 1$, it is not clear whether the $W_{ad}$ is expected to be larger or smaller than $W_{bc}$. The answer depends on the specific $H$-function being used. 
Figure \ref{f:different-h-funcs} displays four different $H$-function families, for both balanced and imbalanced cases. Each of these are created using the same recipe.  Let $F_1$ and $F_2$ be two CDFs, and let $F_{1,2}$ be their convolution CDF. Assuming $x,y \stackrel{ind.}{\sim}$ Unif$(0,1)$, 
\begin{equation}\label{e:HCDF}
H(x, y) = F_{1,2}(F_1^{-1}(x) +F_2^{-1}(y))
\end{equation}
defines an $H$-function with positive association. Figure \ref{f:different-h-funcs} displays the distributions chosen for $F_1$ and $F_2$, and how that choice defines the relative contribution of each input to the result, as well as the shape of the contours of the result.  When $F_1$ and $F_2$ are the same distributional family, and that family has a scale parameter, by the construction of Equation (\ref{e:HCDF}), the \textit{relative} scale of these two distribution drives the $H$-function contours. 
For this reason, we can assume the scale parameter in $F_1 = 1$, and the scale parameter of $F_2$ reflects the relative impact of the second dimension relative to the first, which we call $\rho$. Taking an example from \cite{LeinwandPipiras}, if $F_1$ and $F_2$ are Normal, letting $\rho=\sigma_2^2/\sigma_1^2$, $$H(x,y)= \Phi\left(\frac{1}{\sqrt{1+\rho}}\left(\Phi^{-1}(x) + \sqrt{\rho}\Phi^{-1}(y)\right)\right).$$

In the unweighted network context, we do not need to model edge weights, but instead the probability that an edge is present. Even so, all the properties that make $H$-functions appealing for mapping nodal features to edge weight orderings can also be seen as advantages for mapping nodal features to edge probabilities. In addition to their flexibility, $H$-functions take on values in $[0,1]$ by construction, avoiding incoherent estimates of the kind seen in DCBMs. It must be stated that $H$-functions are not themselves networks; they are functions from $[0,1]\times[0,1] \rightarrow [0, 1].$ They are useful if the connection probability between two nodes in a network with sociability parameters $\in [0, 1]$ can modeled by this kind of function. However, $H$-functions output random variables with support $[0,1]$,
but networks usually have a much narrower range of edge probabilities, so $H$-functions cannot be used to model networks without further adaptation to the unweighted setting.

\section{ACRONYM}
\label{s-unweightedmodel}

 We propose a network generating mechanism where nodes $u, v$ are assigned to communities $i, j$ and provided ``sociability parameters" $\Psi_u$, $\Psi_v \in [0,1]$. Each pair of communities has 3 features (for simplicity of notation, we ignore the $i$ and $j$ community subscripts, but these are assumed present anywhere these parameters appear): $f:[0,1]\longrightarrow [0,1]$ is a monotonic function; $H:[0,1]\times[0,1]\longrightarrow [0,1]$ is an $H$-function; finally, $\sigma$ modulates how much the SC is corrupted by determining the magnitude of errors injected into the underlying probabilities.  
The general model for the edge between nodes $u$ and $v$ given by:
\begin{equation}\label{e:new-model}
A_{uv} \sim \text{Bern}\left(f \left(\Phi\left(\frac{1}{\sqrt{1+\sigma^2}}  \Phi^{-1}(H(\Psi_u, \Psi_v))+ \frac{\sigma}{\sqrt{1+\sigma^2}} \epsilon_{uv}\right)\right)\right).
\end{equation}
There are several important ramifications of this structure. First, assuming $\sigma$ is not too large, the presence or absence of edges are broadly driven by the relevant $H$-function and the node sociability parameters. Second, the impact of SC is mediated depending on the value of $\sigma$. If $\sigma$ is very large, the pattern of edge weights will not depend much on node sociabilities. Third, the choice of $f$ may also impact the underlying probabilities. Two obvious choices for $f$ that generate very different networks are the identity function and a threshold function, that is 
\begin{equation*}
f_t(x) = 
\left\{
    \begin{array}{lr}
        0, &\text{if } x < t\\
        1, &\text{if } x\ge t. 
    \end{array}
\right.
\end{equation*}
       
This paper specifically focuses on the case where $f$ takes the linear form \begin{equation}\label{e:linear-function}
    f(x) = \alpha x + \beta, 
\end{equation} 
where $\beta=\beta_{ij}$ defines the minimal edge presence probability for any potential edge that connects a node in community $i$ with a node in community $j$, and $\alpha=\alpha_{ij}$ represents the width of the interval over the range of probabilities for these edges. To maintain valid probabilities, we require that $0 \le \alpha, \beta, \alpha+\beta \le 1$. Taken together, the full equation is given by: 
\begin{equation}\label{e:new-model-full}
A_{uv} \sim \text{Bern}\left(\alpha\Phi\left(\frac{1}{\sqrt{1+\sigma^2}}  \Phi^{-1}(H(\Psi_u, \Psi_v))+ \frac{\sigma}{\sqrt{1+\sigma^2}} \epsilon_{uv}\right) + \beta \right).
\end{equation}

 The form of (\ref{e:new-model-full}) can create networks with different ranges of probabilities in different pairs of communities, in addition to the potential different patterns of preferences due to different $H$-functions or varying local $\Psi$ values. This structure is reminiscent of different subnetworks of weighted networks having potentially different edge weight distributions, though in this case, the finer variation is only present in the underlying probabilities, not in the observed network's edges, as can be seen in Figure \ref{f:simulated-result}.  The model is thus governed both by community memberships and individual node sociabilities, with different augmented degree correction patterns of connections within each subnetwork induced by pairs of communities. This complicated structure can become somewhat blurred, though, as each edge is still a Bernoulli random variable, albeit each with a different probability parameter.   
Though more general functions may be of interest in future work, we restrict our focus here to linear $f$ functions of the form in (\ref{e:linear-function}), as including a nonlinear $f$ may require the use of tools like logistic regression, and may destructively interfere with particular $H$-functions. 
  
\begin{remark}
When $\alpha=0$ for all subnetworks, the model in Eq. (\ref{e:new-model-full}) reduces to a stochastic blockmodel where the connection probability for all edges within a particular subnetwork is precisely $\beta$. 
The model further reduces to an Erd\H os-R\'enyi (ER) model if all subnetworks share a common $\beta$ value. 
The model can also reduce to an SBM as $\sigma \rightarrow \infty$, as in this case, the edge weight probability becoming $\beta + \frac{\alpha}{2}$. 
These two circumstances are not necessarily identical, though. 
One can imagine repeatedly sampling networks governed by a common set of all parameters including $\epsilon$'s, that is, the only randomness remaining is via the Bernoulli. Conditional on these parameters, when $\alpha=0$, each edge in each network will truly have probability $\beta$ of appearing. When $\sigma \rightarrow \infty$, on the other hand, some edges in  will be much more likely to appear than others, even within the same subnetwork. This paper is not explicitly concerned with repeated network samples, but potential future work interested in, for example, clustering multiple networks with potentially different structures may need to distinguish between these contexts.
Moreover, if the $\Psi$ values are bounded from above or below within a narrower interval than $(0,1)$, it may be difficult to accurately estimate $\alpha$ and $\beta$, as the limited range of edge probabilities may be attributed to those parameters rather the limited range of $\Psi$ values (indeed, model identifiability questions arise in these restricted settings). 
For this reason, it is assumed that $\Psi$ values are allowed near the boundaries of the unit interval. 
\end{remark}

\section{Estimation}
\label{s:unweightedestimation}

Though the generating mechanism of the network is relatively similar to that of weighted networks, the estimation procedures cannot be so easily adapted. 
Edge weights can carry fine-grained information in a weighted network, though observing only a weighted network the distribution of edge weights is unknown and must be estimated. 
An unweighted network, however, is inherently more quantized, and this demands a new approach.
In the weighted setting, \cite{LeinwandPipiras} posits the constraint that within each estimated community, the estimates of the $\Psi$ values are evenly spaced, and then proceeds via the least squares approach. 
As a consequence, letting $\widehat{\Psi}_u^{(j)}$ be the estimate of $\Psi_u$ with respect to community $j$, employing the estimation approach from \cite{LeinwandPipiras} would yield a threshold-type estimate within each subnetwork where $\widehat{A}_{uv} = 1$ anytime  $\Phi^{-1}(\widehat{H}(\widehat{\Psi}_u^{(j)}, \widehat{\Psi}_v^{(i)}))$ exceeds some threshold value $t$, and  $\widehat{A}_{uv} = 0$ otherwise. 
This estimate would not be especially informative, as it does not provide probabilities, but rather just indicates which estimated probabilities exceed the density threshold. 
Furthermore, while the value $\widehat{H}(\widehat{\Psi}_u^{(j)}\widehat{\Psi}_v^{(i)}) \in [0, 1]$, it cannot be directly treated as a probability since the edge probability also depends on the density of the relevant subnetwork. For example, if the subnetwork representing edges between nodes in community $i$ contains 70\% of the potential edges, while the subnetwork between nodes in community $j$ only contains 30\% of the potential edges, for $u_1, u_2 \in i$ and $v_1, v_2 \in j$ it could be the case that $P(A_{u_1 u_2}=1) > P(A_{v_1 v_2}=1)$, even if $\sigma=0$ and $\widehat{H}_{ii}(\widehat{\Psi}_{u_1}^{(i)}\widehat{\Psi}_{u_2}^{(i)}) < \widehat{H}_{jj}(\widehat{\Psi}_{v_1}^{(j)}\widehat{\Psi}_{v_2}^{(j)})$.
 
Considering each edge in the unweighted network as a Bernoulli random variable whose probability is determined by sociability parameters, an $H$-function, $\sigma$, $\alpha$, and $\beta$ as in Eq. (\ref{e:new-model-full}) paves the way for using techniques which seek to maximize the likelihood of the observed network subject to the constraint that parameters are constant within each subnetwork. 
This allows for estimation of these parameters, as well as detection of other distributions of $\Psi$ values beyond the uniform distribution. 
The only restriction for estimation is that there are some $\Psi$ values within each community that are relatively close to 0, and others relatively close to 1. 
Otherwise, the limited probability range may be construed as consequences of  $\alpha$ and $\beta$ parameters, rather than of the $\Psi$ values. 
For this section, we assume the communities have been estimated (correctly), or are known in advance, and estimation of each subnetwork uses only information from that particular subnetwork. Discussion of the community detection preprocessing step is deferred until Section \ref{S:comm-det}.

\subsection{Likelihood based estimation for unweighted networks}
\label{sec:LLH}

For an observed network with adjacency matrix $\mathbf{A}=[A_{uv}]$, the log-likelihood for an estimated matrix $\mathbf{P}=[P_{uv}]$ is given by 
\begin{equation}\label{e:log-lik}
    \sum_{\{u<v\}} A_{uv}\log{P_{uv}} + (1-A_{uv})\log{(1-P_{uv})}. 
\end{equation}
In the context of Eq. (\ref{e:new-model-full}), the network edges are Bernoulli RV's with probability 
\begin{equation}\label{e:edge_prob}
P_{uv} = \alpha\left(\Phi\left(\frac{1}{\sqrt{1+\sigma^2}}  \Phi^{-1}(H_{\rho}(\Psi_u, \Psi_v))+ \frac{\sigma}{\sqrt{1+\sigma^2}} \epsilon_{uv}\right)\right) + \beta.   
\end{equation}
For each subnetwork, our aim is to estimate each of $\alpha, \beta, \sigma,$ the functional form of $H$, the balance of the two arguments of $H$ represented by $\rho$, as well as the $\Psi_u$ and $\Psi_v$ values.  
The ultimate goal is to find parameter estimates whose corresponding plug-in estimates maximize the log-likelihood. 
In principle, we view the $\epsilon_{uv}$ values as $\mathcal{N}(0,1)$ random variables idiosyncratic to this particular network, so we do not wish to incorporate the specific values into our final edge probability estimates, and hence we will soon integrate it out of the $\hat P$ estimates. However, in practice estimating these values can aid in estimating the other parameters. 
As $\epsilon$'s are latent variables with known unconditional distributions, this would seem to be a natural setting for applying the EM algorithm. 
However, given the large number of parameters to be estimated, we instead take a less computationally intensive iterative approach as described in Algorithm \ref{alg:estimateSubnetwork}. 

Based on equation (\ref{e:edge_prob}), throughout our estimation procedure, we will use the plug in estimates and assumed functional form to calculate an intermediate estimate called $\widehat{P}$:
\begin{equation}\label{e:phat}
    \widehat{P}_{uv}=\widehat{\alpha} \Phi\left(\frac{1}{\sqrt{1+\widehat{\sigma}^2}}\Phi^{-1}(\widehat{H}_{\widehat{\rho}}(\widehat{\Psi}_u, \widehat{\Psi}_v)) + \frac{\widehat{\sigma} \check{\epsilon}_{uv}}{\sqrt{1+\widehat{\sigma}^2}}\right) + \widehat{\beta}.
\end{equation}
Here, $\widehat{P}$ includes random $\check{\epsilon}_{uv}$ values centered on the estimates of the $\epsilon$ values. 
Similar to the idea of integrated likelihood, we derive our final edge probability estimates by fixing the estimates of the other parameters and calculating  
\begin{equation}
    \label{eq:exp_eps}
\mathbf{E}_\epsilon\left[\Phi\left(\frac{1}{\sqrt{1+\widehat{\sigma}^2}}\Phi^{-1}(\widehat{H}_{\widehat{\rho}}(\widehat{\Psi}_u, \widehat{\Psi}_v)) + \frac{\widehat{\sigma} \epsilon_{uv}}{\sqrt{1+\widehat{\sigma}^2}}\right)\right]
\end{equation}
under the assumption that $\epsilon \sim \mathcal{N}(0,1)$.
It is shown in \cite{ellison1964two} that if $Z\sim \mathcal{N}(A, B^2)$, then $\mathbf{E}[\Phi(Z)] = \Phi(A/\sqrt{1+B^2}).$  
It follows that Eq. (\ref{eq:exp_eps}) is equal to
$$\Phi\left(\frac{\Phi^{-1}(\widehat{H}_{\widehat{\rho}}(\widehat{\Psi_1}_u, \widehat{\Psi}_v))/\sqrt{1+\widehat{\sigma}^2}}{\sqrt{1+\frac{\widehat{\sigma}^2}{1+\widehat{\sigma}^2}}}\right)=
\Phi\left(\frac{\Phi^{-1}(\widehat{H}_{\widehat{\rho}}(\widehat{\Psi_1}_u, \widehat{\Psi}_v))}{\sqrt{2\widehat{\sigma}^2 + 1}}\right). $$
Note that this expected value includes a $2\hat \sigma^2$ even though (\ref{e:new-model}) has $\hat \sigma^2$ in the denominator.  Based on the derivation above, we set:
$$\widetilde{\Phi}(\widehat{H}_{\widehat{\rho}}(\widehat{\Psi}_u, \widehat{\Psi}_v))=\Phi\left(\frac{\Phi^{-1}(\widehat{H}_{\widehat{\rho}}(\widehat{\Psi}_u, \widehat{\Psi}_v))}{\sqrt{1+2\widehat{\sigma}^2}}\right).$$

This estimate implies that the expected value $\wtp(\widehat{H}_{\widehat{\rho}}(\widehat{\Psi}_u, \widehat{\Psi}_v))$ shrinks somewhat towards .5, with the amount of shrinkage increasing with $\hat\sigma$. In turn, the predicted edge probability shrinks towards the overall density of the subnetwork. 
As we view the $\epsilon$'s as noise values independent of the network parameters, we will use 
$$\Tilde{P}_{uv}= \mathbf{E}_{\epsilon}\left[\widehat{P}_{uv}\right] = \widehat{\alpha}\wtp(\widehat{H}_{\widehat{\rho}}(\widehat{\Psi}_u, \widehat{\Psi}_v)) + \widehat{\beta}$$
as our estimate of the generating process rather than $\widehat{P}$. In Sections \ref{s:simulation}, \ref{s:congress}, and \ref{s:MR-data} we treat this $\Tilde{P}$ as our final estimate.

\subsection{Parameter Estimation Details}

The proposed estimation approach described at a high level in the pseudocode in Appendix \ref{sec:pseudo}.
Note that our approach models each subnetwork independently of all the others. This allows for estimating different $\Psi$ values for the same node in different regions of the network. 
This can be beneficial in terms of flexibility, but if each node only has a single $\Psi$ value everywhere, this approach will be over-parameterized and not able to borrow strength across the whole network. 
For each subnetwork, we must estimate the form of the $H$-function, the $\alpha, \beta, \rho, \sigma$ and $\Psi$ values. 
In practice, we restrict our potential $H$-functions to four families, those that are seen in Figure \ref{f:different-h-funcs}, which, from left to right, we refer to as ``normal," ``concave," ``convex," and ``linear," based on the shape of their contours. 
The iterative estimation procedure described in Algorithm \ref{alg:estimateSubnetwork} is initiated and run for each of these $H$-functions for 5 iterations. For computational efficiency, we choose only the family which is performing best (according to likelihood) after these initial 5 iterations and run Algorithm \ref{alg:estimateSubnetwork} for 95 additional cycles under that regime. 
Depending on context and available computational resources, alternative schemes such as running all families for the same number of iterations, or to some stopping criterion, may be preferable.

Given a network clustering $\{\mathfrak{c}_i\}_{i=1}^K$ (see Section \ref{S:comm-det}) and parameter initializations (see Appendix \ref{sec:Init} for details on how we initialize the parameters herein), our iterative procedure proceeds as follows.  
Denote our current parameter estimates for the $\mathfrak{c}_i$-$\mathfrak{c}_j$ subnetwork via 
$\wha,\whb,\whs,\whr,\whP,\whe,\che$ and the best achieved parameter estimates via 
$\what,\whbt,\whst,\whrt,\whPt$, then 
\begin{itemize}
\item[i.] For each vertex pair $\{u \in \mathfrak{c}_i,v \in \mathfrak{c}_j\}$ in the subnetwork compute
$$\widehat{P}_{uv} = \wha\Phi\left(\frac{1}{\sqrt{1+\whs}}\Phi^{-1}(H_{\whr}(\whP_u, \whP_v)) + \frac{\whs \che_{uv}}{\sqrt{1+\whs}}\right) + \whb$$
and the current log-likelihood
$\ell=  \sum_{u}\sum_{v} A_{uv}\log \widehat{P}_{uv} + (1-A_{uv})\log(1-\widehat{P}_{uv})$.
Note that in each iteration, we do not optimize the other parameters using $\widehat{\epsilon}_{uv}$ directly, but instead draw a random value $\check{\epsilon}_{uv}$ centered at $\widehat{\epsilon}_{uv}$ for each entry (see Step ii. below). This is discussed further below in Remark \ref{rem:eps}.
\item[ii.]  Use $\widehat{P}_{uv}$ to update the values of $\whe_{uv}$, the best estimate of $\epsilon_{uv}$ for each edge given the observed edge value $A_{uv}$ and our other estimates. 
The update is done via
\begin{equation}
\label{e:eps}
    \whe_{uv} = \argmax {\epsilon} \quad \phi(\epsilon_{uv})\left((\breve{\breve{P}}_{uv})^{A_{uv}}(1-\breve{\breve{P}}_{uv})^{(1-A_{uv})}\right),
\end{equation}
where $\phi$ is the standard normal density, and $\breve{\breve{P}}_{uv}$ is as in (\ref{e:phat}) though here depends on $\epsilon_{uv}$ via
$$\breve{\breve{P}}_{uv} = \wha\Phi\left(\frac{1}{\sqrt{1+\whs}}\Phi^{-1}(H_{\whr}(\whP_u, \whP_v)) + \frac{\whs \epsilon_{uv}}{\sqrt{1+\whs}}\right) + \whb$$
For edges where $A_{uv}=1$, the optimization proceeds via first taking the log of Eq. (\ref{e:eps}) and then computing the following,
where $\phi$ is the standard normal density and $\Phi$ the standard normal CDF)
\begin{align*}
&\frac{d}{d\epsilon}\left(\log\left(\phi(\epsilon_{uv})\right) + \log\left(\widehat{\alpha}\Phi\left(\frac{1}{\sqrt{1+\widehat{\sigma}^2}}\Phi^{-1}(\widehat{H}_{\widehat{\rho}}(\widehat{\Psi}_u, \widehat{\Psi}_v)) + \frac{\widehat{\sigma} \epsilon_{uv}}{\sqrt{1+\widehat{\sigma}^2}}\right) +\widehat{\beta}\right)\right)\\
&= \frac{\phi^\prime(\epsilon_{uv})}{\phi(\epsilon_{uv})} + 
\left(\frac{\widehat{\alpha}\phi\left(\frac{1}{\sqrt{1+\widehat{\sigma}^2}}\Phi^{-1}(H_{\widehat{\rho}}(\widehat{\Psi}_u, \widehat{\Psi}_v)) + \frac{\widehat{\sigma} \epsilon_{uv}}{\sqrt{1+\widehat{\sigma}^2}}\right)}{\widehat{\alpha}\Phi\left(\frac{1}{\sqrt{1+\widehat{\sigma}^2}}\Phi^{-1}(\widehat{H}_{\widehat{\rho}}(\widehat{\Psi}_u, \widehat{\Psi}_v)) + \frac{\widehat{\sigma} \epsilon_{uv}}{\sqrt{1+\widehat{\sigma}^2}}\right) +\widehat{\beta}} \right) \frac{\widehat{\sigma}}{\sqrt{1+\widehat{\sigma}^2}}. 
\end{align*}
Rather than directly finding the critical point which may be computationally costly (i.e., where this derivative is $0$), we seek equivalently to find (noting that the derivative of a standard normal pdf $\phi(X)$ is given by $-\phi(x)*x$)
$$\widehat{\epsilon}_{uv}=\argmin{\epsilon}\left(\epsilon - \left[ \frac{\widehat{\alpha}\phi\left(\frac{1}{\sqrt{1+\widehat{\sigma}^2}}\Phi^{-1}(\widehat{H}_{\widehat{\rho}}(\widehat{\Psi}_u, \widehat{\Psi}_v)) + \frac{\widehat{\sigma} \epsilon}{\sqrt{1+\widehat{\sigma}^2}}\right) \frac{\widehat{\sigma}}{\sqrt{1+\widehat{\sigma}^2}}}{\widehat{\alpha}\Phi\left(\frac{1}{\sqrt{1+\widehat{\sigma}^2}}\Phi^{-1}(\widehat{H}_{\widehat{\rho}}(\widehat{\Psi}_u, \widehat{\Psi}_v)) + \frac{\widehat{\sigma} \epsilon}{\sqrt{1+\widehat{\sigma}}}\right) +\widehat{\beta}}\right]\right)^2, $$
which is more amenable to computational optimization. 
The corresponding result for edges where $A_{uv}=0$ is 
$$\widehat{\epsilon}_{uv}=\argmin{\epsilon}\left(\epsilon + \left[ \frac{\widehat{\alpha}\phi\left(\frac{1}{\sqrt{1+\widehat{\sigma}^2}}\Phi^{-1}(\widehat{H}_{\widehat{\rho}}(\widehat{\Psi}_u, \widehat{\Psi}_v)) + \frac{\widehat{\sigma} \epsilon}{\sqrt{1+\widehat{\sigma}^2}}\right) \frac{\widehat{\sigma}}{\sqrt{1+\widehat{\sigma}^2}}}{1-\widehat{\alpha}\Phi\left(\frac{1}{\sqrt{1+\widehat{\sigma}^2}}\Phi^{-1}(\widehat{H}_{\widehat{\rho}}(\widehat{\Psi}_u, \widehat{\Psi}_v)) + \frac{\widehat{\sigma} \epsilon}{\sqrt{1+\widehat{\sigma}^2}}\right) -\widehat{\beta}}\right]\right)^2. $$
Once $\whe_{uv}$ is estimated, update  $\che_{u,v}\sim$Norm($\,\whe_{uv},1)$.
\item[iii.] Update the following three parameter sets (a., b., and c. below) in a uniformly random order.  
Note that when the subnetwork is within a single community, we must constrain $\widehat{\rho} = 1$, and $\{\widehat{\Psi}_u\}=\{\widehat{\Psi}_v\}$. In those cases, we randomly choose whether to simultaneously update  $\wha,\whb,\whs$ first or all of the $\whP_u$ values simultaneously first.  
\begin{itemize}
    \item[a.]\textbf{Updating $\widehat{\alpha}, \widehat{\beta}, \widehat{\rho},$ and $\widehat{\sigma}$:}  To avoid potentially incoherent estimates, we constrain $\wha$ and $\whb$ to values slightly larger than 0 and slightly less than 1; in practice, constraining them to be in $(.001, .999)$ seems to be effective.
    We further constrain $\whr$ and $\widehat{\sigma}$ to be between $.1$ and $10$, as this represents sufficiently extreme bounds such that anything beyond these bounds will not meaningfully impact estimates (recall though that $\widehat{\rho}$ is constrained to be exactly 1 for symmetric subnetworks within a single community). 
    Here, we update via
    $$\{\whatc,\whbtc,\whrtc,\whstc\}\leftarrow \text{argmax}_{\wha,\whb,\whr,\whs}\sum_{u}\sum_{v} A_{uv}\log \widehat{P}_{uv} + (1-A_{uv})\log(1-\widehat{P}_{uv}).
    $$
  In practice, we perform this optimization via the \texttt{nlminb} function in \RR.
    Compute
    $$\widehat{P}^{(\text{cand})}_{uv} = \whatc\Phi\left(\frac{1}{\sqrt{1+\whstc}}\Phi^{-1}(H_{\whrtc}(\whP_u, \whP_v)) + \frac{\whstc \che_{uv}}{\sqrt{1+\whstc}}\right) + \whbtc$$
    If 
    $$
    \sum_{u}\sum_{v} A_{uv}\log \widehat{P}_{uv} + (1-A_{uv})\log(1-\widehat{P}_{uv})<\sum_{u}\sum_{v} A_{uv}\log \widehat{P}^{(\text{cand})}_{uv} + (1-A_{uv})\log(1-\widehat{P}^{(\text{cand})}_{uv})
    $$
    then update
    $\{\wha,\whb,\whr,\whs\}\leftarrow \{\whatc,\whbtc,\whrtc,\whstc\},$
    else do not update the parameters.
 Note that the observed log-likelihood is not guaranteed to improve at every step, as the random choices of $\che$ may be deleterious. 
    \item[b.]\textbf{Updating all $\widehat{\Psi}_u$ for $u\in\mathfrak{c}_i$:} Note that each node has \textit{K} different $\Psi$ values, one for each subnetwork in which it appears. 
    Here, we are updating the $\widehat{\Psi}^{ij}_u=\widehat{\Psi}_u$ corresponding to the subnetwork between $\mathfrak{c}_i$ and $\mathfrak{c}_j$.
    We drop the superscript below to ease notation.
    Use the current $\whP_v$ estimates and the current estimates for $\widehat{\alpha}, \widehat{\beta}, \widehat{\rho},$ and $\widehat{\sigma}$.
    For each $u\in\mathfrak{c}_i$, compute
    $$\whPtc_u\leftarrow \text{argmax}_{\whP_u}\sum_{v} A_{uv}\log \widehat{P}_{uv} + (1-A_{uv})\log(1-\widehat{P}_{uv})
    $$
    In practice, we perform this optimization via the \texttt{optim} function in \RR $ $  and constraining the values to be between $\Phi(-4)$ and $\Phi(4)$.
    Compute then
    $$\widehat{P}^{(\text{cand})}_{uv} = \wha\Phi\left(\frac{1}{\sqrt{1+\whs}}\Phi^{-1}(H_{\whr}(\whPtc_u, \whP_v)) + \frac{\whst \che_{uv}}{\sqrt{1+\whs}}\right) + \whb.$$
    If  
    $$
    \sum_{u}\sum_{v} A_{uv}\log \widehat{P}_{uv} + (1-A_{uv})\log(1-\widehat{P}_{uv})<\sum_{u}\sum_{v} A_{uv}\log \widehat{P}^{(\text{cand})}_{uv} + (1-A_{uv})\log(1-\widehat{P}^{(\text{cand})}_{uv})
    $$
    then update $\{\whP_u\}_{u\in\mathfrak{c}_i}\leftarrow \{\whPtc_u\}_{u\in\mathfrak{c}_i}$, else do not update the parameters.
    Consequently, even if the model would be improved by changing certain individual $\widehat{\Psi}_u$ values, if the aggregate change does not lead to an overall improvement in likelihood, these individual changes will not be made. 
    This is to avoid over-fitting and provide algorithmic stability by avoiding constantly changing values to chase random fluctuations in $\check{\epsilon}$.  
    \item[c.]\textbf{Updating all $\widehat{\Psi}_v$ for $v\in\mathfrak{c}_j$:} 
    Update the collection $\{\widehat{\Psi}^{ij}_v\}_{v\in\mathfrak{c}_j}$ via an identical routine to step b. in which the $\{\whP^{ij}_u\}_{u\in\mathfrak{c}_i}$ were updated.
\end{itemize}  
\item[iv.] Again compute the current log-likelihood
$\ell=  \sum_{u}\sum_{v} A_{uv}\log \widehat{P}_{uv} + (1-A_{uv})\log(1-\widehat{P}_{uv})$.
If this log-likelihood is higher than the previous best log-likelihood, then update 
$\{\what,\whbt,\whst,\whrt,\whPt\}\leftarrow \{\wha,\whb,\whs,\whr,\whP\}$.  
\end{itemize}
After the final iteration of updating parameters, compute $\widehat{P}$ using $\{\what,\whbt,\whst,\whrt,\whPt\}$ and $\check{\epsilon}_{uv}=0$. 
Calculate $\widehat{\epsilon}_{uv}$ using this new $\widehat{P}$, and re-estimate only $\widehat{\sigma}$ (see Remark \ref{rem:eps} below) using $\widehat{\epsilon}_{uv}$ rather than $\check{\epsilon}_{uv}$. 
Finally, calculate our final $\widehat{P}, \tilde{P}$ and likelihood, and return those values along with the estimated parameters.  Note that when the density of a subnetwork exceeds 50\%, the same process described above is instead applied to estimate the probability an edge is \textit{not} present, using the input $1 - A_{uv}$. Once the estimation procedure is complete, the final estimated quantities of interest can be backsolved using the estimates on the inverted data. 

\begin{remark}
\label{rem:eps}
The reason to use the random $\check{\epsilon}$'s instead of the optimized $\widehat{\epsilon}$'s is to avoid over-fitting, based on the current potentially poor parameter estimates. For one thing, the value of $\widehat{\epsilon}$ updated in Step ii of the algorithm is negative for every 0-valued edge, and positive for every 1-valued edge. Additionally the collective distribution of these conditionally likelihood maximizing $\epsilon$ values does not represent their unconditional distribution, the standard Normal. Randomly sampling around $\widehat{\epsilon}_{uv}$ appears to improve estimation performance in practice, likely by allowing the estimates to escape local minima stemming from inaccurate initialized parameters. 

We now address the reasoning for the re-estimation of $\sigma$.
Re-estimation yields practical gains, as throughout estimation we wish to take advantage of the random $\epsilon$ values to avoid getting stuck in a local minimum. 
If we do not re-estimate, our algorithm can inappropriately drive $\sigma$ estimates to 0. 
We re-estimate at the end, when all our other estimates are set, as then we can take those values to get a better estimate of all $\epsilon$ values, which will allow us to get a better estimate of $\sigma$.
\end{remark}

\begin{remark}
Note that the naive initialization estimates provided in Appendix \ref{sec:Init} may lead us to a local optimum, whereas if we knew the truth we would be able to achieve a better result. 
For example, see the 5th column in the bottom row of Figure \ref{f:Psi-image}, where the true $\Psi$ values take a concave shape, but the estimates appear to be approximately linear. Given the flexibility of the problem, though, we do not currently have a proposed method of avoiding this issue in the most general case.
\end{remark}

\section{Community detection} 
\label{S:comm-det}
Existing methods -- developed either for weighted networks with augmented degree correction, or for unweighted networks -- do not appear to work well for detecting communities to use as input in ACRONYM. 
We note here that due to the network generating mechanism in (\ref{e:new-model-full}), the extremal signal eigenvalues of the network itself have both positive and negative sign.
Additionally, the first eigenvector is usually highly correlated with the degree of each node, not the community structure. By row-normalizing the matrix, all the eigenvectors useful for community detection are shifted to the front, though the corresponding eigenvalues only decrease in absolute value, with some being positive and some negative.
The eigenvectors associated with the normalized network appear to lie along a manifold, and can be projected onto the surface of a hypersphere. The process is detailed below. 

 We begin by row-normalizing the $A$ matrix by taking $N_{uv} = (A_{uv} - \overline{A_{u\bullet}})/\text{sd}(A_{u\bullet})$ where $\overline{A_{u\bullet}}$ (resp., $\text{sd}(A_{u\bullet})$) represents the mean (resp., standard deviation) of the row in $A$ corresponding to node $u$. 
 This is similar to the methodology in \cite{lei2016goodness,bickel2016hypothesis}, in which a similar statistic was used to test the goodness of fit of a proposed SBM model.  
 As noted there, it is reasonable to expect the $N$ matrix to be approximately a generalized Wigner ensemble, and the machinery of random matrix theory (see, for example, \cite{erdHos2012rigidity,erdHos2012spectral,erdHos2013spectral}) can be brought to bear.
 Here, we use the heuristics of \cite{chatterjee2015matrix,zhu2006automatic} to estimate the number of signal dimensions by looking for elbows in the eigenvalue scree plot.
Note that since this $N$ matrix is not necessarily symmetric, we plot the absolute values of the real parts of the eigenvalues of $N$ and look for the elbow, which we will denote via $\tilde{d}$. 
In other network contexts, this $\tilde{d}$ might be our choice of $\widehat{k}$, but here, it is merely an intermediate step. 
For these $\tilde{d}$ eigenvectors of $N$, each row of this $n \times \tilde{d}$ matrix is further projected onto the surface of a hypersphere as done when clustering DCBMs in \cite{lyzinski2014perfect} and more recently in \cite{passino2022spectral}; note that the projected $v$-th row is denoted via $\harpoon{u}_v$. 
We then compute the cosine distance (i.e., $1-\text{cos}(\harpoon{u}_v, \harpoon{u}_w)$ for vertices $v$ and $w$) and proceed to cluster the rows/columns of this distance matrix using standard hierarchical clustering.
For the simulated example in Section \ref{s:simulation}, using \texttt{single-linkage} hierarchical clustering and cutting off the dendrogram at the greatest height gap yields the correct results. 
For the congressional dataset as well as the mouse retina dataset, this same approach results in several singleton communities. However, we can recover a few communities with several members by using \texttt{ward.d} clustering to build the dendrogram in the hclust function in R instead of single-linkage clustering. This is discussed further in Section \ref{s:congress}.

\begin{figure}[t!]
\begin{center}
\includegraphics[width=0.8\linewidth]{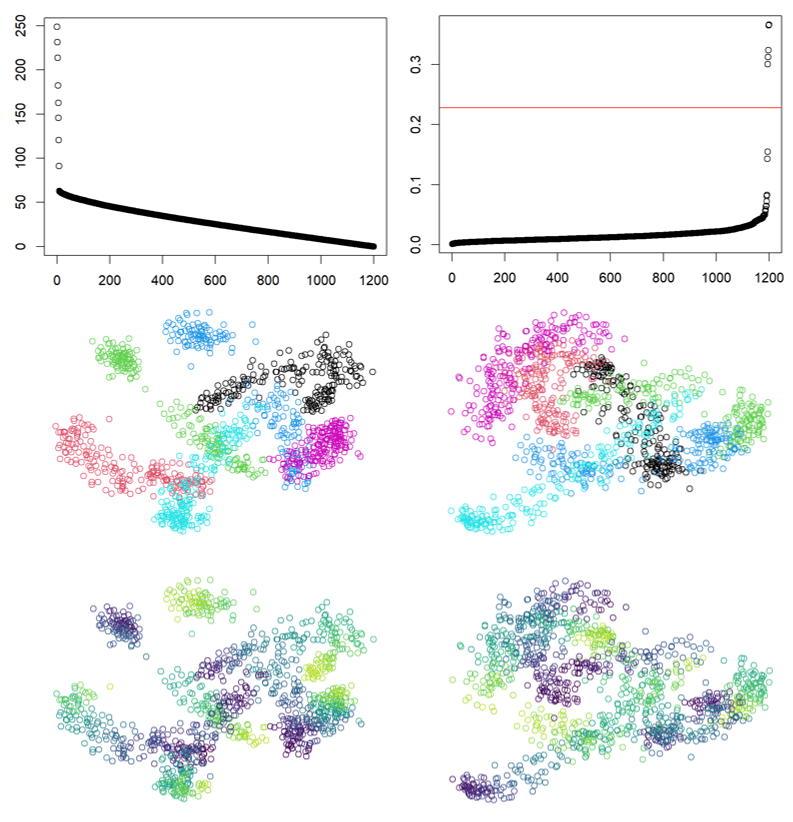}
\caption{Top left: Absolute values of eigenvectors of the normalized matrix for the simulated dataset. The first 8 values are separated from the bulk. Top right: Dendrogram heights of single-linkage hierarchical clustering for the $\{\harpoon{u}\}$ values. Middle left: the first and second entries of the ${\harpoon{u}}$ vectors for each node, colored by community membership. Middle right: the third and fourth entries of the ${\harpoon{u}}$ vectors for each node, colored by community membership. Bottom left: the first and second entries of the ${\harpoon{u}}$ vectors for each node, colored by their $\Psi$ values. Bottom right: the third and fourth entries of the ${\harpoon{u}}$ vectors for each node, colored by their $\Psi$ values.}
\label{f:simulated-comm-det}
\end{center}
\end{figure}

In the eigenvectors of the normalized matrix, often nodes from each community seem to arrange along a manifold, which is unsurprising in light of \cite{whiteley2022statistical,rubin2020manifold}.
Because of this structure, representations of nodes in the same community are not necessarily all close to one another, though they point in a similar direction, as some nodes are farther from the origin in certain dimensions and closer in others. 
That motivates using cosine similarity, but even then, one can imagine that nodes in the same community with wildly different $\Psi$ values will have representations along the same manifold, but at different ends of it. That is in fact what we see in Figure \ref{f:simulated-comm-det}, demonstrating the community detection approach to the simulated network of Section \ref{s:simulation}.
For those communities where the $\Psi$ values have a gap, we see this same gap in the eigenvectors of the normalized matrix.  
At this stage, though, we are only interested in community detection, not $\Psi$ estimation. 
To that end, since these communities do not surround a single centroid, single-linkage clustering is suited to finding these communities. 
However, even in our simulated case with gaps, the missing $\Psi$ values are between .4 and .6 or .55 and .8, with many remaining nodes on either side. One can imagine a situation where most community members have very small $\Psi$ values but a few have very large $\Psi$ values. Single linkage clustering might see this as two different communities if there is no bridge to connect them.

\section{Simulation}
\label{s:simulation}

\begin{figure}[t!]
\includegraphics[width=\linewidth]{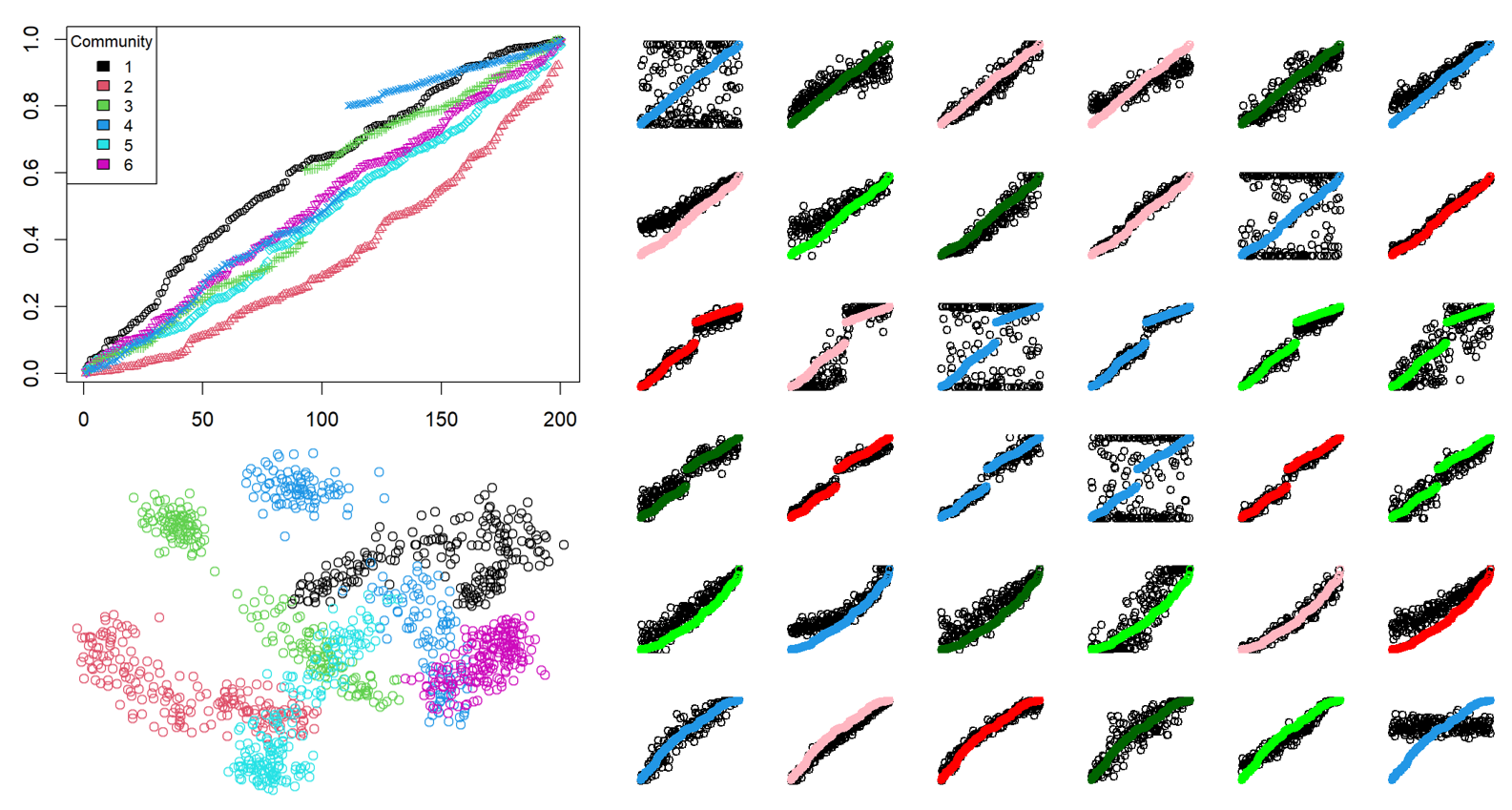}
\caption{Top left: The sorted $\Psi$ values for all 200 nodes in each community for the simulation outlined in Section \ref{s:simulation}. 
Bottom left: the first and second entries of the ${\harpoon{u}}$ vectors for each node, colored by their community membership. 
In two dimensions, the community structure on the hypersphere is not clear. However, the gaps in the $\Psi$ values for the 3rd and 4th communities -- in green and dark blue, respectively -- appear in this representation as well. 
Right: The $\Psi$ value for each node is estimated 6 times, one for each community in the network. In each row, the estimates for $\Psi$ for that row's community are shown in black, while the true values are shown in color. The bottom row depicts 6 different estimates of the $\Psi$ values for nodes in community 1, which have a concave shape, using the community in the corresponding column. True values shown in blue indicate $\rho=1$, that is, that the two communities have equal influence on the edge probability. True values shown in pink indicate the depicted community had somewhat less influence on the edge probability than the other community (at least 1/4 as influential as the other community), while those in red have far less influence (less than 1/4 as influential as the other community). True values shown in light green are somewhat more influential than the other community on edge probability (at most 4 times as influential), while those in dark green have far more influence (at least 4 times as influential as the other community). }
\label{f:Psi-image}
\end{figure}

We generate a network to illustrate the strengths and limitations of the proposed estimation framework. 
The network consists of 1200 nodes, split into 6 communities of 200 nodes each. All nodes are assigned a $\Psi$ value, but different communities have different distributions of $\Psi$ values. The first community draws $\Psi$'s from a $Beta(1.5, .9)$ distribution, while the second community draws from a $Beta(.8, 1.4)$. These distributions were chosen to have different shapes.
In particular, when plotting the sorted $\Psi$ values (as in Figure \ref{f:Psi-image}), the first community's curve looks concave, while the second's looks convex. 
At the same time, both distributions are not so concentrated in one region.
Practically, this yields $\Psi$ values non-uniformly distributed over $[0,1]$. 
The third community's $\Psi$ values are distributed uniformly, except no $\Psi$ values appear in the range $(.4, .6)$. The fourth community similarly has a gap in the range $(.55, .8)$. The $\Psi$ values for the fifth and sixth communities are drawn from a $\text{Unif}(0,1)$. The actual values are shown in the top left plot of Figure \ref{f:Psi-image}.

\begin{figure}[t!]
\begin{center}
\includegraphics[width=0.7\linewidth]{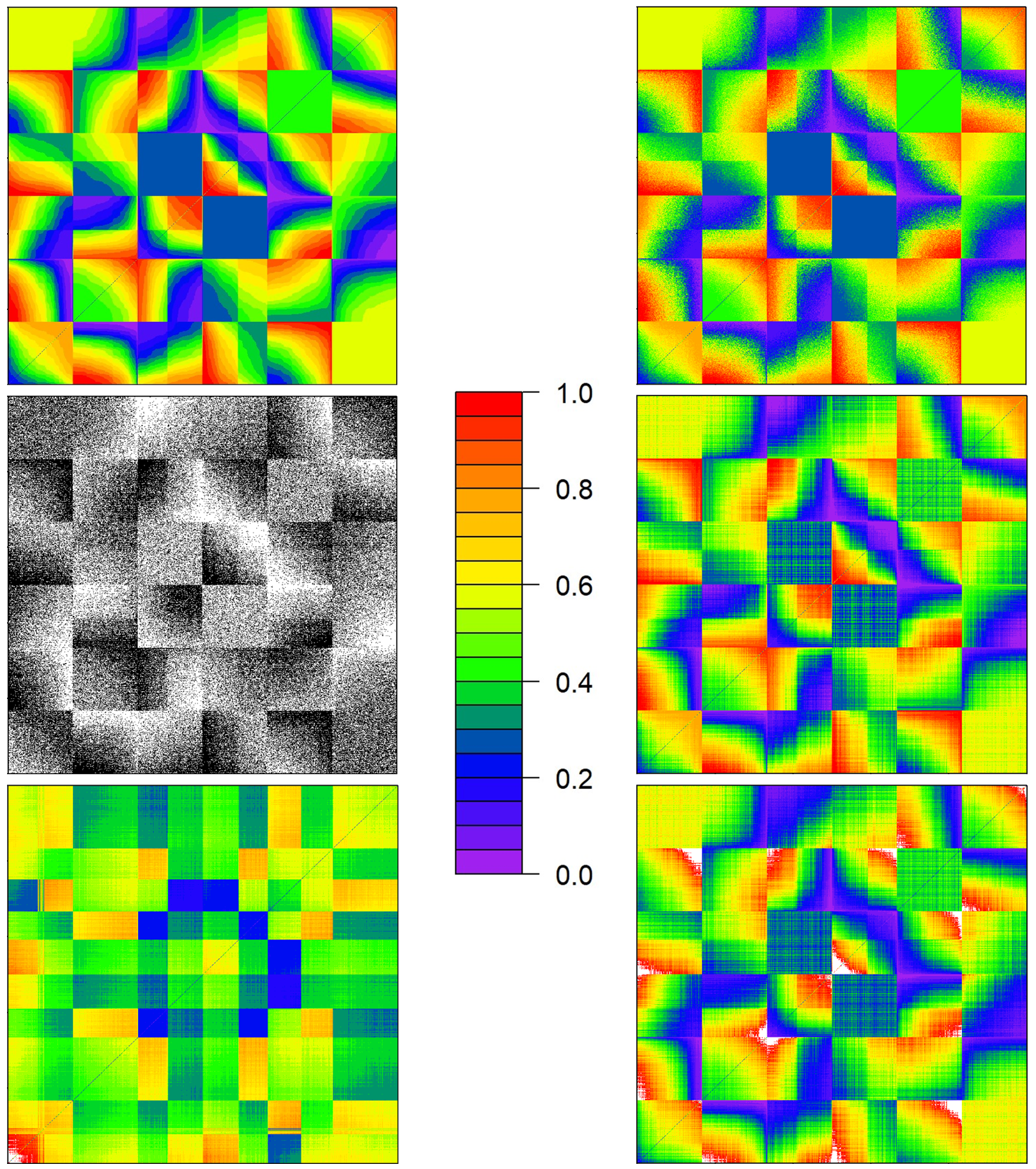}
\caption{Top left: Matrix displaying equation ($\ref{e:edge_prob}$)  with all $\epsilon$ values set to 0 for the simulated network. Top right: Matrix displaying equation ($\ref{e:edge_prob}$)  with drawn $\epsilon$ values for the simulated network. Middle left: The observed simulated network. Middle right: The estimated $\tilde{P}$ matrix for the observed network. Bottom left: The estimated probabilities using a DCBM fit using communities detected using the \texttt{nett} package. Bottom right: The estimated probabilities using a PABM fit using the true communities.}
\label{f:simulated-result}
\end{center}
\end{figure}

Each pair of communities is given a particular $\alpha, \beta, \rho,$ $H$-functional form, and type of association. The $\sigma$ values for within community edges are set to .3, and set to .4 for between community edges. In the top left plot of Figure \ref{f:simulated-result}, the underlying probabilities are shown if $\epsilon$ values are all set to 0. That is, it illustrates the matrix of Equation (\ref{e:edge_prob}) when $\epsilon_{uv}=0$ for all $u, v$ pairs. 
We call this matrix $P^*$. 
The subnetwork within the first community is displayed on the bottom left, and the communities increase while moving up and to the right. 
Furthermore, the nodes are ordered by their $\Psi$ value within each community. 
The top right plot in Figure \ref{f:simulated-result} shows the true underlying probabilities, here denoted $P$, or Equation (\ref{e:edge_prob}) drawing random standard Normal $\epsilon$'s. Notice this plot reflects broadly the same pattern of probabilities as in $P^*$, though is visually noisier. 
The exceptions are the subnetworks between communities 3 and 4, between communities 1 and 6, and within community, as these are ER graphs, where $\alpha=0$ and hence not dependent on $\epsilon$. 

Looking at these two plots, the ``gaps" in the $\Psi$ values for communities 3 and 4 are visible throughout those rows and columns, as there is discontinuity in an otherwise similar pattern. 
Moreover, the ``normal," ``concave," ``convex," and ``linear" $H$-function patterns are discernible, as are the positive, negative, and Simpson associations, including a negative association within community 4. 
Differing $\rho$ values are somewhat harder to detect, but can be observed in the subnetwork between communities 1 and 2, or the subnetwork between communities 5 and 6. 
These $\rho$ values are 1 by construction in the diagonal blocks of within community subnetworks. 
The disparity in $\sigma$ values are not immediately obvious using visual inspection, though $\alpha$ and $\beta$ values can be somewhat inferred by the limited range of colors in e.g. the subnetwork within community 1, or the subnetwork between communities 2 and 6. 
All things considered, visual inspection of these underlying probability matrices reveals a lot of information about the network, where clear patterns arising from the choice of $\Psi$ values, $H$-functions, and other network parameters.

Unfortunately, we often only observe the unordered adjacency matrix rather than these structured $P$'s.
In the middle left plot of Figure \ref{f:simulated-result}, the observed network is shown, with black areas representing edges, and white areas representing non-edges. 
Note that this graph is ordered via the ordering of the $P^*$ matrix.
As expected, the observed network appears to mute some of the visually discernible features discussed previously. 
In the middle right plot of Figure \ref{f:simulated-result}, our ACRONYM estimate, denoted $\tilde{P}$, is shown, and we see that these features are recovered reasonably well, including the ``gaps" in the third and fourth communities. Note that the nodal ordering is the same as in the top left plot.  
Some nodes appear to be out of place, because of incorrect $\Psi$ estimation, where in this case, we see ``plaid" patterns in certain subnetworks, as these errors occur at the node level within each subnetwork, not at the edge level as in the plot of $P_{uv}$. This is most obvious in those subnetworks which are truly ER graphs, as the random Bernoulli draws yield some variation in the observed edges, and hence in our estimated probabilities (see the left panel of Figure \ref{f:Psi-image} where we show the $\Psi$ estimates for each subnetwork).
Fortunately, there appears to be little variation in the estimated probabilities within those regions, and they appear to be approximately reflect the true underlying probability. 
This is one potential drawback of estimating all subnetworks separately; some nodes may have their $\Psi$ values correctly estimated in most of the network, but incorrectly with respect to a particular other community. If the $\Psi$ values do not vary with respect to the other node's community membership, as in this network, one may be able to estimate $\Psi$'s more robustly using information from across the network. 
Better $\Psi$ estimates would presumably improve other parameter estimates, as in these ER subnetworks, which may estimate $\alpha$ to be closer to 0 if the $\Psi$ values are estimated to vary more.
This is made more obvious when looking at the estimates of the $\Psi$ values, as shown in the right plot of \ref{f:Psi-image}. In general, it looks easier to estimate $\Psi$'s for nodes in communities 5 and 6, followed by communities 3 and 4, with communities 1 and 2 being the most difficult. This may be due to the initialization approach of the $\Psi$ estimates, which assumes uniformly distributed $\Psi$ values. The only cases where these estimates deviate dramatically from the true estimates are in the ER subnetworks. 
Quantifying the error rates in these cases is ongoing work, as the rates likely depend non-trivially on the choice of parameters and $H$-functions.  

 In this case, it is known how the network is generated, and that the community detection approach, at the very least, correctly identified the intended community structure. Additionally, the estimated $\Psi$ values are broadly, if imperfectly, aligned with the true values. While Figure \ref{f:simulated-result} appears to indicate that the estimation procedure is effective, mere visual inspection of a network this complicated may be misleading \cite{peixoto_2023}. 
 When actually looking at the likelihood of this estimate against the likelihoods (where the log-likelihood is denoted $\ell$) of the two matrices at the top of Figure \ref{f:simulated-result}, we have 
 $\ell(P|A)>\ell(\widetilde P|A)>\ell(P^*|A)$.
 Numeric summaries of the computed likelihoods and MSE (when estimating $P^*$) for our method and the DCBM and PABM are shown in Table \ref{t:sim-results}.
 Note that our clustering algorithm correctly identified the true $6$ communities here, hence ACRONYM using the true community memberships.
From the table, we see that our estimation procedure is an improvement over the DCBM and PABM alternatives.
Moreover, this example demonstrates the limitations of existing methods (here from the \texttt{randnet} \cite{randnet} and \texttt{nett} \cite{nett} R package) when estimating the number of communities using off-the-shelf methods. 
The method of \cite{le2022estimating} (resp., \cite{bickelwang}) chooses 5 (resp., 16) as the optimal number of communities, and we estimate 5 (resp., 16) communities using spectral clustering, regularized spectral clustering \cite{rohe2011spectral, amini2013pseudo,lei2015consistency, le2017concentration} and regularized spherical spectral clustering \cite{qin2013regularized}.
When estimating the DCBM and PABM edge probabilities given these communities, some edges have an estimated probability greater than 1; these edges are excluded from the MSE estimates. 
As the simulated network contains subnetworks with negative associations, and others with Simpson associations, the assumptions for DCBM networks should not hold, and so it is not be a surprise that clustering mechanisms designed around these assumptions are not effective. 
Interestingly, using the true 6 communities and fitting a DCBM on these communities does not yield any incoherent probability estimates. However, the edge probability estimation suffers from the same struggle with non-positive associations, so the likelihood for that model is much worse than $\tilde{P}$. 
On the other hand, the PABM should be able to incorporate these local patterns. 
Sparse subspace clustering implemented in R fails on this dataset, so to simplify, we use the true community structure to fit a PABM. Surprisingly, using the true community structure and fitting a PABM will yield incoherent estimates on more edges than any other model. 

In Table \ref{t:sim-results}, the likelihoods are calculated only at edges with a valid probability estimate.
Note that the unconstrained PABM with the true 6 communities has a superficially better likelihood performance than ACRONYM, but this is measured on 12,000 fewer edges. 
By taking all predicted values in the PABM that exceed 1 and setting them to .999, we can see how this truncated PABM (which does not have the correct expected degree over the whole network) performs compared to our model on all edges in the network. With this modification, the negative log-likelihood of the truncated PABM on the whole network is worse than ACRONYM's. 
ACRONYM's MSE to $P^*$ is also impressive. In both this metric and on likelihood, models with more communities do better as expected, as there are more parameters available to fit. Again, the PABM performs reasonably well, as it can capture different association directions across different parts of the network, but the issues of invalid probabilities and a fixed functional form are detrimental.      

\begin{table}[t!] \label{t:sim-results}
    \centering
    \begin{tabular}{|p{5cm}|p{1.8cm}|p{3cm}|p{3cm}| p{3cm}|}
    
    \hline
        \textbf{Probability model} &\textbf{Communities} &  \textbf{Edges with valid estimated probabilities} & \textbf{$-\ell(A|\cdot)$ on edges with valid estimated probabilities} &\textbf{MSE from $P^*$ on edges with valid estimated probabilities} \\ \hline
        $P^*$ & 6 (true) & 719,400 &  420,582 & 0  \\ \hline
        $P$ & 6 (true) & 719,400 &  410,569 & .00553 \\ \hline
        ACRONYM $\tilde{P}$ & 6 (true) & 719,400 &  416,945 & .00222 \\ \hline
        DCBM & 6 (true) & 719,400 &  481,696 & .04088  \\ \hline
        PABM & 6 (true) & 707,183 &    414,249 & .00381   \\ \hline
        PABM (Truncated at .999) & 6 (true)  & 719,400 &  423,321 & .00386 \\ \hline
        DCBM with spectral clustering with regularization (nett)   & 5 & 718,518 &   467,841 & .02996  \\ \hline
        DCBM with regularized spectral clustering (randnet) & 5 & 718,960 &  468,640 & .03027 \\ \hline
        DCBM with regularized spherical spectral clustering (randnet)& 5 & 719,399 &  469,444 & .03162 \\ \hline
        DCBM with spectral clustering with regularization (nett)  & 16 & 719,277 &  433,656 & .00882  \\ \hline
        DCBM with regularized spectral clustering (randnet) & 16 & 719,285 &  434,501 & .00933  \\ \hline
        DCBM with regularized spherical spectral clustering (randnet) & 16 & 719,302 &  435,680 & .01016 \\ \hline
    \end{tabular}\vspace{.05in}
    \caption{Table describing different models fit to the simulated dataset. The first 2 rows are generated along with the observed network.  Our estimate $\tilde{P}$ recovers the true communities. For the DCBM and PABM models in the next 2 rows, the true communities are assumed known,  while the remaining rows use alternative methods of community detection. The third column counts the number of potential edges each model estimates to have probabilities which lie in $[0,1]$, out of a total of 719,400 potential edges. The fourth column measures the negative log-likelihood only where the probability estimates are valid, that is, on each model's valid potential edges. The fifth column measures the average entrywise squared distance from $P^*$ of these models, again only where the probability estimates are valid.}  
\end{table}

\section{Congressional Twitter Dataset}\label{s:congress}
\begin{figure}[t!]
\includegraphics[width=\linewidth]{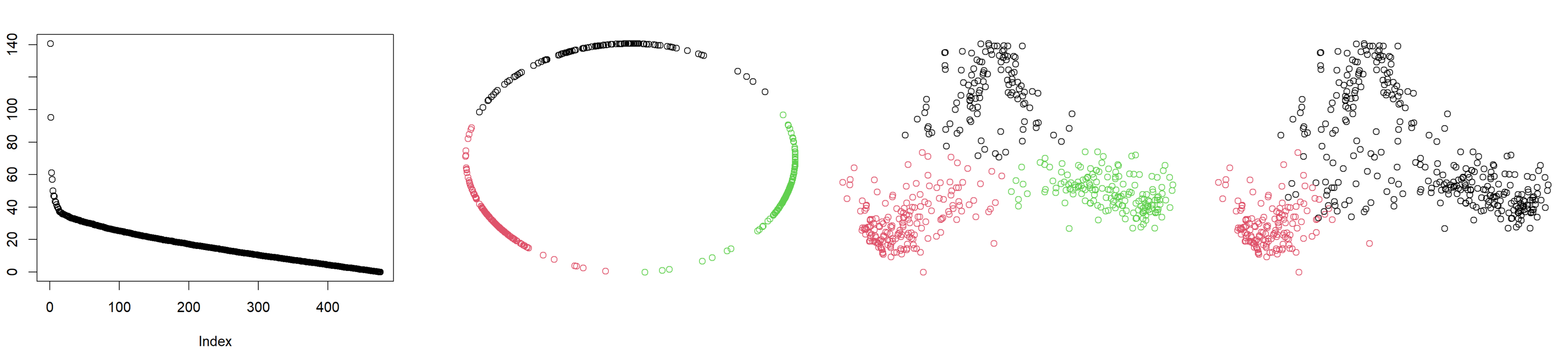}
\caption{From left to right: the real eigenvalues of the normalized matrix derived from the congressional Twitter network. The first and second entries of the $\harpoon{u}$ vectors for each node when $\tilde{d} = 2$, colored by the communities recovered when $\tilde{d} = 2$. The first and second entries of the $\harpoon{u}$ vectors for each node when $\tilde{d} = 4$, colored by the communities recovered when $\tilde{d} = 2$. The first and second entries 0f the $\harpoon{u}$ vectors for each node when $\tilde{d} = 4$, colored by the communities recovered when $\tilde{d} = 4$.}
\label{f:congress-comm-det}
\end{figure}

As described earlier, we also examine a network where each senator or congressperson serving from February 9, 2022 to June 9, 2022 who issued at least 100 tweets in that period is a node, and there is an edge between a pair of nodes if either actor interacted with the other on Twitter (now known as ``X")
during that period. 
Unlike in the simulated example, in this case, the true underlying probabilities are unknown, so there is a risk of over-fitting, in part because ACRONYM includes more inputs than alternative models even given the same community structures. One option to avoid attributing the success of our model to over-fitting is to use K-fold cross-validation to examine our results on held out edges. 
For the estimation procedure, we ignore those hidden edges, and maximize the likelihood based on what is available (i.e., not held out) to the model. 
To overcome the issue with spectral community detection method not working when values are missing, we first estimate the number of communities using our clustering method (here given by 3); see Figure \ref{f:congress-comm-det}.
With these communities, we fit ACRONYM and  DCBM and PABM (using the \texttt{nett} and \texttt{randnet} packages) to the whole network.
We also perform 10-fold cross-validation for ACRONYM, where we note that the number of communities in the full network was used as a baseline for selecting the number of communities in the CV steps, leading to potential information leakage in the CV procedure.
We do community detection 10 times (once for each full network less the hidden fold). 
To do so, for each fold, we normalize the network (leaving the held-out edges as missing) and calculate the row means and standard deviations on the edges from the remaining folds. Then, use the DINEOF (Data Interpolating Empirical Orthogonal Function) \cite{beckers2003eof, taylor2013sensitivity} function from the \texttt{sinkr} repository \cite{sinkR}, to estimate the remaining values in the normalized network. 
Community detection continues on that imputed network. 
ACRONYM is then estimated on the network (with held-out edges not imputed).
Results are summarized in Table \ref{t:congress-results}.

In contrast to the simulated network, the ``true" communities of the congressional network are unknown (and the notion ill-defined). 
For the full network, the community detection approach was attempted with $\tilde{d}=2$, as shown in Figure \ref{f:congress-comm-det}. In this case, there is not quite such a clear separation of eigenvalues, and looking at the 2- dimensional normalized $\harpoon{u}$ vectors indicates there may be other potential clusters that differ from the 3 recovered communities.  However, looking at these same communities plotted against the first  2 entries of $\harpoon{u}$ vectors had we set $\tilde{d}=4$, it looks like these 3 communities are appropriate (especially compared with the clustering on $\tilde{d}=4$, which only recovers 2 communities). 
Of the 92 senators in the dataset, all but 1 are assigned to the first community, which also includes 25 congresspeople. This is the smallest community, with community 2 containing 175 members (including Senator Bernie Sanders), and community 3 containing 184 members, none of whom are senators.
Note that for each of the 10 folds for cross-validation, $\tilde{d}$ is set to 2 as the eigenvalues look similar to the full network. 
Due to this information leakage, cross-validation results for this dataset may be somewhat overoptimistic.
Furthermore, each fold may recover a different number of communities. 

The results are similar to those of the simulated dataset, and they are displayed in Table \ref{t:congress-results}. Sparse subspace clustering implemented in \texttt{R} fails on this data, so we use the recovered communities on the whole network from our approach to estimate DCBM and PABM parameters. 
Both the Bethe-Hessian method of \cite{le2022estimating} and likelihood-ratio approaches of \cite{bickelwang} choose 7 as the number of communities, so DCBMs are estimated using different approaches 
from the \texttt{nett} and \texttt{randnet} packages with $K=7$ as well. 
All models besides ACRONYM estimate incoherent probabilities exceeding 1. PABM appears to perform best according to negative log-likelihood, but when capping the probabilities to .999 and including all edges, it is still inferior to ACRONYM.  The cross-validated ACRONYM estimated probabilities outperform all other models, surprisingly including the ACRONYM probabilities estimated using the whole observed network. ACRONYM's advantage when compared to DCBM with $K=7$ is less pronounced in this dataset. This may be because of the lower density, or less clear non-positive associations, but also potentially because 7 communities allows for more severe over-fitting to the observed data. Still, that the alternatives estimate nonsensical link probabilities is a severe flaw. 

\begin{table}[b!] 
    \centering
    \begin{tabular}{|p{4.2cm}|p{1.8cm}|p{3cm}|p{4cm}|}
    
    \hline
        \textbf{Probability model} &\textbf{Communities} &  \textbf{Potential edges with valid probabilities} & \textbf{Negative log-likelihood on potential edges with valid probabilities only} \\ \hline
       ACRONYM $\tilde{P}$	&	3	&	 112,575 	&	 24,159 	\\ \hline
ACRONYM 10 Fold CV $\tilde{P}$	&	NA	&	 112,575 	&	 24,097 	\\ \hline
DCBM	&	3	&	 112,328 	&	 25,925 	\\ \hline
PABM	&	3	&	 112,257 	&	 24,061 	\\ \hline
PABM Truncated at .999	&	3	&	 112,575 	&	 24,344 	\\ \hline
DCBM with \texttt{nett} spectral clustering (with regularization) 	&	7	&	 112,135 	&	 25,044 	\\ \hline
DCBM with \texttt{randnet} regularized spectral clustering	&	7	&	 112,136 	&	 25,197 	\\ \hline
DCBM with \texttt{randnet} regularized spherical spectral clustering	&	7	&	 112,209 	&	 25,074 	\\ \hline

    \end{tabular}\vspace{.05in}

 \caption{Table displaying the performance of different models fit to the Congressional Twitter dataset.}  \label{t:congress-results}
\end{table}
 
\section{Mouse Retina Dataset}\label{s:MR-data}
As another demonstration of the ACRONYM approach to a different domain, we fit several models to the dataset consisting of the structural map of  ``the mouse inner plexiform layer--the
main computational neuropil region in the mammalian retina" from \cite{helmstaedter2013connectomic}, which can be found at \url{https://github.com/ericmjonas/circuitdata/tree/master/mouseretina}. 
The original dataset contains 1123 nodes, but 47 of those nodes have no edges, so they are removed from the dataset. 
Furthermore, though there are multiple contacts between different cells, for the purposes of this paper, we only analyze the unweighted adjacency matrix, which includes 1076 nodes and 90,811 edges out of a possible 578,350 edges, giving a density of about 15.7\%. 

In this case, we run cross validation using only 3 folds for ACRONYM, so there is quite a bit of missing data to be estimated in each fold. Most of the estimation mirrored the approach described in Section \ref{s:congress}. However in this case, the number $\tilde{d}$ of eigenvectors to keep in each fold during community detection was chosen by visual inspection rather than based on a prior analysis, and was picked to be 5 for all folds. 
In 2 of the 3 folds, 5 communities were recovered, while in the third fold, 4 communities were recovered. 
Following estimation in each fold, the estimates of the hidden values from each fold are collated to create the 3 fold CV estimate. In addition to the cross-validation analysis, community detection and estimation was conducted on the full dataset, and those estimated communities from the full network were utilized to assess the performance of the DCBM and PABM, along with estimates from the \texttt{nett} and \texttt{randnet} packages.

\begin{figure}[t!]
\begin{center}
\includegraphics[width=0.8\linewidth]{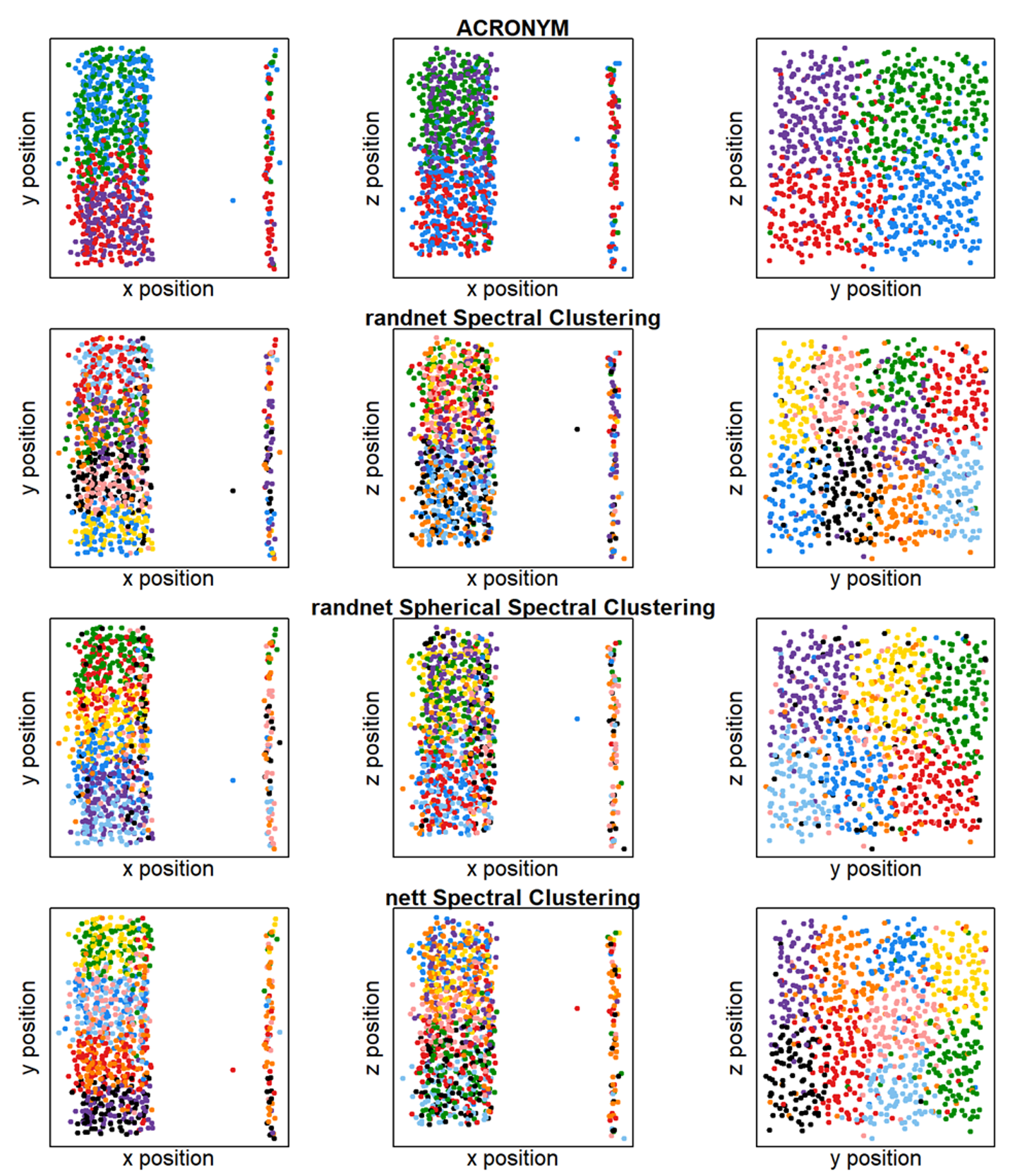}
\caption{Community detection results of our method and those from the \texttt{nett} and \texttt{randnet} packages, plotted against the locations of the soma of the cells. In all cases, community detection results are most clearly mapped to the y-z plane, though our method finds fewer communities than the alternatives.}
\end{center}
\label{f:retina-commdet}
\end{figure}

Figure \ref{f:retina-commdet} displays the recovered communities using our method on the fully observed dataset as well as those from the \texttt{nett} and \texttt{randnet} packages plotted against the locations of the soma of the cells in the given dataset; 
note that these are the derived $(x,y,z)$ coordinates that were provided for each soma giving the location in the physical retina.
These results show relatively clear spatial community boundaries in the y-z axis. Our approach recovers fewer communities than the alternatives, but all seem to be capturing similar information. Though the x-axis does not appear informative here, the dataset includes cell type information, and the coarse type of cell does broadly depend on the position in the x dimension.       
\begin{table}[b!] 
    \centering
    \begin{tabular}{|p{4cm}|p{1.8cm}|p{3cm}|p{4cm}|}
    
    \hline
        \textbf{Probability model} &\textbf{Communities} &  \textbf{Potential edges with valid probabilities} & \textbf{Negative log-likelihood on potential edges with valid probabilities only} \\ \hline
ACRONYM $\tilde{P}$	&	4	&	 578,350 	&	 142,346 	\\ \hline
ACRONYM 3 Fold CV $\tilde{P}$	&	5, 5, 4	&	 578,350 	&	 148,237 	\\ \hline
DCBM	&	4	&	 569,519 	&	 169,878 	\\ \hline
PABM	&	4	&	 567,673 	&	 147,535 	\\ \hline
PABM truncated	&	4	&	 578,350 	&	 154,991 	\\ \hline
\texttt{nett} spectral clustering (with regularization) 	&	9	&	 568,453 	&	 160,350 	\\ \hline
\texttt{randnet} regularized spectral clustering	&	9	&	 568,561 	&	 159,783 	\\ \hline
\texttt{randnet} regularized spherical spectral clustering	&	9	&	 569,506 	&	 157,694 	\\ \hline
\end{tabular}\vspace{.05in}
\caption{Table displaying the performance of different models fit to the Mouse Retina dataset.}\label{t:retina-results}
\end{table}
Table \ref{t:retina-results} summarizes the results, which follow the same pattern as before. 
All other models estimate incoherent probabilities. ACRONYM fitted on the fully observed dataset obtains a better likelihood than any comparable alternative (i.e., using the same number of edges). 
Even with more than 10,000 additional valid edges and observing only 2/3 of the data in the CV, ACRONYM nearly outperforms PABM. 
Partially observed ACRONYM handily outperforms any DCBM model, as well as the truncated version of PABM. Though it is not immediately discernible using this result, beyond the other benefits discussed earlier, there is hope that ACRONYM can yield better predictions by using fewer and larger communities, leaving more local information to draw upon for those missing edges.

\begin{figure}[t!]
\includegraphics[width=\linewidth]{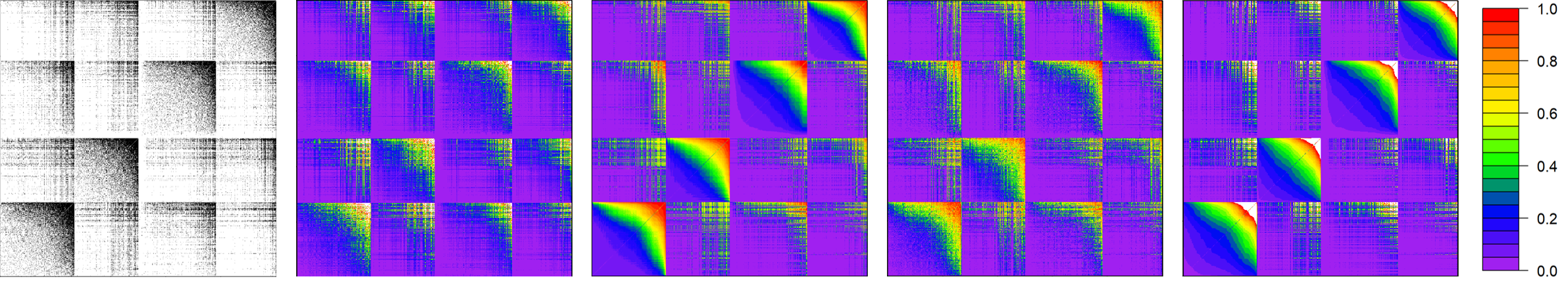}
\caption{From left to right: the underlying dataset; estimates from DCBM using the \texttt{nett} R package; estimates from ACRONYM using the full dataset; predictions from ACRONYM using cross-validation; estimates from PABM using communities recovered using our approach. All images in this plot are reordered first by estimated community using our approach on the  fully observed network, then by within estimated community degree.}
\label{f:retina-estimates}
\end{figure}

Figure \ref{f:retina-estimates} shows the dataset, the estimated probabilities using ACRONYM on the fully observed dataset, the predicted probabilities using cross-validated ACRONYM, and the estimated probabilities using PABM and DCBM. All images in Figure \ref{f:retina-estimates} are reordered first by estimated community using our approach on the fully observed dataset, then by within estimated community degree. There are a couple of salient features in this plot. First, in the estimates from PABM and DCBM, incoherent estimates (those greater than 1) are colored white. They are concentrated in regions of high density. Second, because of this ordering matching the community structure used to fit the PABM, the PABM image has perfect ordering within each community, since only local degree matters. 
This is not the case for the DCBM plot because global sociability may be different than local sociability. Perfect within community ordering in terms of degree may not always occur using ACRONYM because estimated $\Psi$ values do not need share the same ordering as local degree. The likelihood maximizing $\Psi$ value for each node within each subnetwork does not depend only on the number of local edges, but also on the estimated $\Psi$'s of those adjacent nodes. When stitching ACRONYM's predictions from different folds together, inconsistencies across estimates naturally yield noisier looking and worse performing predictions than when the full dataset is observed.     

\begin{figure}[b!]
\includegraphics[width=\linewidth]{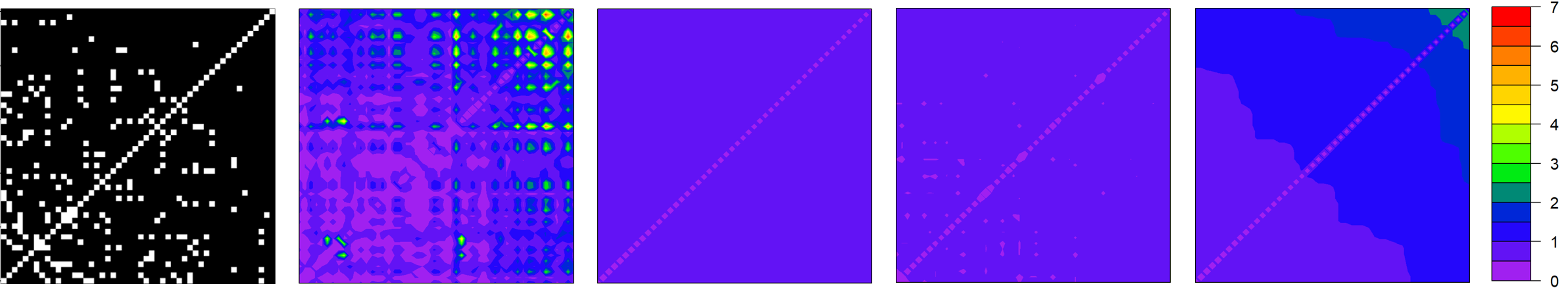}
\caption{Displaying the subnetwork consisting of the 50 nodes in the 4th recovered community with the greatest within community degree, and zooming in on the corresponding regions from Figure \ref{f:retina-estimates}. }
\label{f:zoomed-in}
\end{figure}

In Figure \ref{f:zoomed-in}, we zoom in on the upper right hand corner of Figure \ref{f:retina-estimates}. The 4th community our method recovers has 235 nodes, so we show the induced subnetwork created by taking the top 50 of these 235 nodes, as sorted by within community degree, along with the estimates of this subnetwork from Figure \ref{f:retina-estimates}. Since these are within community estimates, we are looking at 1225 different edges, each shown above and below the diagonal, as well as a diagonal which is constrained to be 0.  In most other figures in this paper, we constrain the values in figures to be between 0 and 1, with values exceeding 1 depicted in white. In this figure, we extend the range of the plots and make their scale less granular in order to portray both the magnitude and the pattern of pervasiveness of incoherent over-prediction from the other models. It becomes easier to see that DCBM here has extreme overestimates, even exceeding 6.5, but because it also incorporates global information, the pattern is harder to discern locally. Conversely, while PABM doesn't overestimate any single edge to the same degree, 814 of the potential 1225 edges in the depicted region are estimated incoherently, compared to ``only" 376 edges for the DCBM. The PABM uses only local information to form estimates, but regularly overshoots reasonable estimates. Our estimates for the full dataset look entirely static in this figure, excluding the diagonal. This is collateral damage from reducing granularity, as the true predictions here range from .697 to .973.  The predictions from the cross-validated version of our approach display a bit more variation, and in fact range from .179 to .982. While the lower end of these estimates is quite poor in this context, the missing data may have impacted the community detection as well as the estimation. The impact is limited to relatively few edges, as only 47 edges are predicted to have a probability less than .5. As with the fuller dataset, our cross-validated approach still empirically looks preferable to the DCBM and PABM approaches.

\section{Conclusion}\label{s:conclusion}

We have introduced a new model for generating networks with different characteristics than others in the literature, though the generating mechanism can nonetheless be summarized using few parameters relative to the number of edges. We have demonstrated two real-world networks where this approach compares favorably to established alternatives, and built synthetic networks to illustrate how these advantages might accrue. This model may provide a new understanding of how certain systems operate, providing novel insights in domains where network data is available. One interesting domain is delving further into neuroscience, including looking for different hierarchical structures from coarse to fine resolutions, similar to \cite{lyzinski2016community, lei2020consistency}. As communities defined by our proposed framework differ from the usual definition of communities, the results may
provide a different perspective on brain organization.

There are many potential avenues to extend this work. The proposed approach may benefit from further work rigorously establishing a set of  $H$-functions which may serve as a basis for the class of $H$-functions. Even with the complexity of the simulated network, based on results in \cite{leinwandrec}, it is worth considering how to deal with $H$-functions which may be generated as compositions of other $H$-functions. Furthermore, extending this model to allow nodes to have mixed-memberships may be beneficial. Joint estimation of model parameters across subnetworks may improve estimation robustness, potentially by taking a Bayesian approach to estimation. Connecting this model to existing approaches like Random Dot Product Graphs  \cite{sussman2012consistent, RDPG, athreya2021estimation}, similar to \cite{koo2023popularity},  may serve as a means of deriving error bounds on community detection and estimation.   

\begin{figure}[t!]
\includegraphics[width=\linewidth]{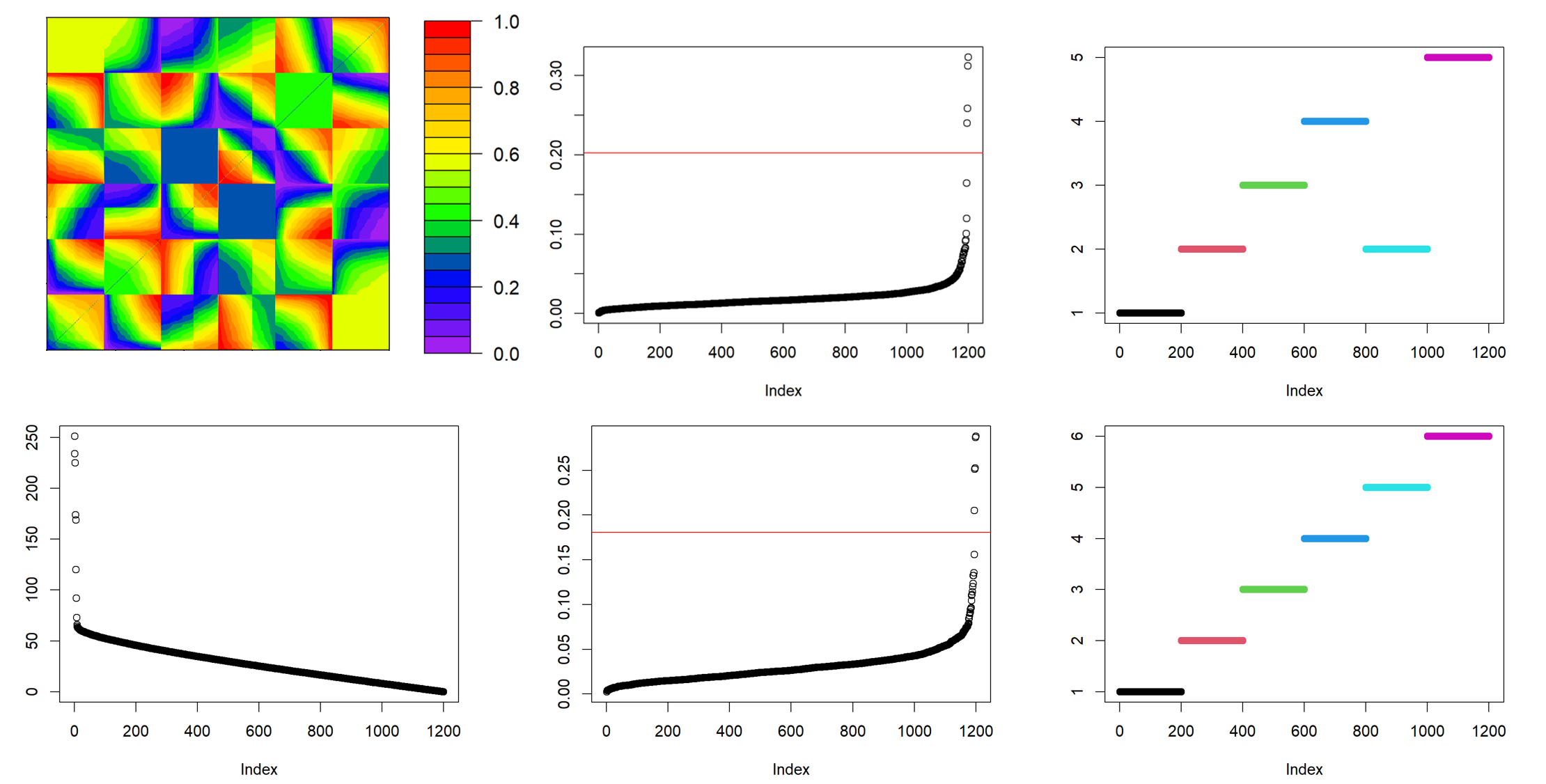}
\caption{Top left: $P^*$ for network where community detection approach fails. Bottom left: Absolute values of eigenvalues of normalized matrix of network. Top center: Dendrogram heights of single-linkage hierarchical clustering for the $\{\harpoon{u}\}$ values in this network when $\tilde{d}=8$. Bottom center: Dendrogram heights of single-linkage hierarchical clustering for the $\{\harpoon{u}\}$ values in this network when $\tilde{d}=9$. Top right: Recovered communities when $\tilde{d}=8$ (height) do not reflect the true underlying structure (color). Bottom right: Using $\tilde{d}=9$ recovers the true communities.}
\label{f:commdet-fail}
\end{figure}

The simulation in Section \ref{s:simulation} was explicitly designed to demonstrate features of the proposed model.
However, different model conditions would yield different results.
For example, see Figure \ref{f:commdet-fail} for an example of how the community detection can fail by changing the association in the subnetwork between communities 1 and 2 from negative to positive, and changing the association in the subnetwork between communities 2 and 5 from Simpson to positive.
By using only the first 8 eigenvectors of the normalized matrix, only 5 communities are detected, merging the second and fifth communities. The community information is still available, but it is embedded elsewhere. For this alternative simulation, if the first 9 eigenvectors of the normalized matrix are included, the true communities are recovered. This points to a potential resolution limit issue, which is reminiscent of the problem of choosing the number of communities $K$ in general network settings.  
In future work, a deeper dive into the spectral properties of the proposed approach to community detection might yield better insight into its asymptotic and finite sample performance. Additional theoretical consideration of community detection and edge probability estimation is required to determine the reliability of these procedures. 

\bibliography{NSFrefs}

\newpage
\appendix

\section{Parameter Initialization}
\label{sec:Init}

Herein we outline a procedure for initializing the parameters of our model for our sequential optimization procedure.  Note that these parameters are initialized independently within each subnetwork, and that the implicit dependence on the subnetwork is assumed (though not explicitly indexed, i.e., in subnetwork $t$, we initialize $\widehat\beta=\widehat\beta_t$ where the dependence on $t$ is understood).
Given community estimate $\{\mathfrak{c}_i\}_{i=1}^K$ that partition the network, we proceed as follows.
\begin{itemize}
\item{\textbf{Initializing $\widehat{\Psi}$:}} One simple approach is to rank the row nodes and column nodes by their local degree (that is, their degree only within this particular subnetwork, only with respect to the nodes in the other community), and divide by the number of nodes in their community plus 1. Formally, letting $D_{\mathfrak{c}_j}(u)$ be the local degree of node $u \in \mathfrak{c}_i$ with respect to community $\mathfrak{c}_j$, and letting $|\mathfrak{c}_i|$ be the number of nodes in community $\mathfrak{c}_i$, $$\widehat{\Psi}_u^{(\mathfrak{c}_j)} = \frac{\sum_{u' \in \mathfrak{c}_i}\mathds{1}\{D_{\mathfrak{c}_j}(u')\le D_{\mathfrak{c}_j}(u)\}}{|\mathfrak{c}_i|+1}.$$ This ensures initial estimates will not be at the boundary, which may cause estimates of 0 or 1, and also represents the expected value of the order statistic of the corresponding rank when observing $|\mathfrak{c}_i|$ uniform random variables.

\item{\textbf{Initializing $\widehat{\beta}$:}} We initialize $\widehat{\beta}$ to be the minimum proportion of existing edges in this subnetwork over the nodes in either of these communities, i.e., as
$$\widehat{\beta}=\min{\left(\frac{\min{(D_{\mathfrak{c}_j}(u))}}{|\mathfrak{c}_i|}, \frac{\min{(D_{\mathfrak{c}_i}(v))}}{|\mathfrak{c}_j|}\right)}$$ 
\item{\textbf{Initializing $\widehat{\alpha}$:}} We initialize $\widehat\alpha$ based on the expected density of the subnetwork. 
Assuming $\Psi$ values are uniformly distributed, the output of any $H$-function fed a random pair of nodes will also be uniformly distributed, so the expected density is $$\mathbf{E}\left[\frac{1}{|\mathfrak{c}_i|\cdot|\mathfrak{c}_j|}\sum_{u \in \mathfrak{c}_i, v \in \mathfrak{c}_j}A_{uv}\right] = \frac{\alpha}{2} +\beta.$$ 
With this naive assumption on $\Psi$, we take the observed density of the subnetwork and solve for $\widehat{\alpha}$ using our estimated $\widehat{\beta}$, note that to avoid pathologies we initially limit $\widehat{\alpha}$ to be between .01 and .99. \\
\item{\textbf{Initializing $\widehat{\rho}$:}} We initialize $\widehat{\rho}=1$, yielding equal balance in our $H$-function between the 2 communities. \\
\item{\textbf{Initializing $\widehat{\sigma}$:}} 
We initialize $\widehat{\sigma}=1$, to provide initial balance between sociability and error. \\
\item{\textbf{Initializing $\epsilon$:}} The initial values of all $\widehat{\epsilon}_{uv}$ and $\check{\epsilon}_{uv}$  are set to be 0. 
\end{itemize}

\newpage 
\section{Algorithmic Pseudocode}
\label{sec:pseudo}

\begin{algorithm}[h!]
\caption{Full Network Estimation Procedure}
\label{alg:fullNetworkEstimation}
\begin{algorithmic}
\Require 
\State $A$: the network
\State $\{\, \fc_i \}_{i=1}^K$: estimated communities/subnetworks
\Ensure
 \For{$i$ from 1 to $K$}
\For{$j$ from i to $K$}
     \State $concave \gets estimateSubnetwork(A^{(\,\fc_i,\fc_j \,)}, \,  concave\,\, \text{H-}function,\, numberIterations = 5, \, completedIterations = 0)$
       \State $convex \gets estimateSubnetwork(A^{(\,\fc_i,\fc_j \,)}, \,  convex\,\, \text{H-}function,\, numberIterations = 5, \, completedIterations = 0)   $ 
       \State $normal \gets estimateSubnetwork(
       A^{(\,\fc_i,\fc_j \,)}, \,  normal\,\, \text{H-}function,\, numberIterations = 5, \, completedIterations = 0)$
       \State $linear \gets estimateSubnetwork(
       A^{(\,\fc_i,\fc_j \,)}, \,  linear
       \,\, \text{H-}function,
       \, numberIterations = 5, \, completedIterations = 0)$
      \State $optimal\,H \gets \argmin{\{\text{concave, convex, normal, linear}\} } (Likelihood)$
      \State $subnetworkEstimate \gets estimateSubnetwork(A^{(\fc_i,\fc_j)}, \, optimal\,H, 100, 5, bestModel(\widehat{\alpha},\widehat{\beta},\widehat{\sigma}, \widehat{\rho}, \{\widehat{\Psi}\}) )$   (Algorithm \ref{alg:estimateSubnetwork})
    \State $\tilde{P}^{(\fc_i,\fc_j) }  \gets subnetworkEstimate(\tilde{P})$ 
    \If{$j > i$}
    \State    $\tilde{P}^{(\fc_j,\fc_i) } \gets (\tilde{P}^{(\fc_i,\fc_j) })^T$
    \EndIf
\EndFor
\EndFor
\State \Return $\tilde{P}$
\end{algorithmic}
\end{algorithm}

\newpage

\begin{algorithm}
\caption{estimateSubnetwork}
\label{alg:estimateSubnetwork}

\begin{algorithmic}
\Require 
\\
\noindent\textbf{\textit{Mandatory inputs:}}
\State $A^{(\, \widehat{i}, \widehat{j} \,)}$: Matrix representing the relevant subnetwork
\State $functionalForm$: Functional form of $H$-function
\State $numberIterations$: Total number of iterations
\State $completedIterations$: Number of already completed iterations

\noindent\textbf{\textit{Optional inputs:}}
\State $\widehat{\alpha}, \widehat{\beta}, \widehat{\sigma}, \widehat{\rho}, \widehat{\Psi}$: ACRONYM estimated parameters

\Ensure
\State \textbf{1. Initialization}
\State bestLikelihood = $\infty$ 
\If{Not all optional inputs have been provided}
\State $\{\widehat{\alpha}, \widehat{\beta}, \widehat{\sigma}, \widehat{\rho}, \widehat{\Psi}\} \gets \text{initialize parameters as in Appendix \ref{sec:Init}}$  
\EndIf

\State $\{\widehat{\epsilon}_{uv}\} \gets 0$ 

\State 
$\text{run Algorithm \ref{alg:updateP} to update the Probabilities, Likelihoods, and Best Parameters}$

\State \textbf{2. Iterative estimation}
\For{$iteration$ from $completedIterations+1$ to $numberIterations$}
\State Calculate $\{\widehat{\epsilon}_{uv}\}$ 
\State sample $\{\che_{u,v}\} \sim \mathcal{N}(\{\widehat{\epsilon}_{uv}\}, 1)$
\State run the following in randomized order:
\State\quad\quad - update $\Psi_u()$ via Algorithm \ref{a:update-Psi};
\State\quad\quad - update $\Psi_v()$ via Algorithm \ref{a:update-Psi};
\State\quad\quad - update the local parameters via Algorithm \ref{a:updateLocalparams}
\State $\text{run Algorithm \ref{alg:updateP} to update the Probabilities, Likelihoods, and Best Parameters}$

\EndFor
\State \textbf{3. Final update of $\widehat{\sigma}^2$  using} $\widehat{\epsilon}$ 

\State $\{\widehat{\alpha}, \widehat{\beta}, \widehat{\sigma}, \widehat{\rho}, \widehat{\Psi}\} \gets bestParameters$

\State $\{\che_{u,v}\} \gets 0$
\State Calculate $\{\widehat{\epsilon}_{uv}\}$
\State $\text{run Algorithm \ref{alg:updateP} to update the Probabilities, Likelihoods, and Best Parameters}$

\State $\{Estimate_{uv}\} \gets   \widehat{\alpha}\left(\Phi_1\left(\frac{1}{\sqrt{1+\widehat{var}^2}}  \Phi_1^{-1}(H_{\widehat{\rho}}(\widehat{\Psi}_v, \widehat{\Psi}_v))+ \frac{\widehat{var}}{\sqrt{1+\widehat{var}^2}} \widehat{\epsilon}_{uv}\right)\right) + \widehat{\beta}$\\
\State$\widehat{\sigma} \gets \argmin{var}  -(\sum_{u}\sum_{v} A_{uv}\log{Estimate_{uv}} + (1-A_{uv})\log{(1-Estimate_{uv})})$  \\
\State $\widehat{P} \gets Estimate$\\
 $\tilde{P} \gets \widehat{\alpha}\Tilde{\Phi}(\widehat{H}(\widehat{\Psi}_u, \widehat{\Psi}_v)) + \widehat{\beta}$\\
\State \Return $\{\widehat{\alpha}, \widehat{\beta}, \widehat{\sigma}, \widehat{\rho}, \widehat{\Psi}, \widehat{P}, \Tilde{P}, Likelihood, functionalForm\}$ 
\end{algorithmic}
\end{algorithm}

\begin{algorithm}[h!]
    \caption{update Probabilities, Likelihoods, and Best Parameters Procedure}
    \begin{algorithmic}
        \State  $\widehat{P} \gets \widehat{\alpha} \Phi\left(\frac{1}{\sqrt{1+\widehat{\sigma}^2}}\Phi^{-1}(\widehat{H}(\widehat{\Psi}_u, \widehat{\Psi}_v)) + \frac{\widehat{\sigma} \che_{u,v}}{\sqrt{1+\widehat{\sigma}^2}}\right) + \widehat{\beta}$ \\
\State  $Likelihood \gets  -(\sum_{u}\sum_{v} A_{uv}\widehat{P}_{uv} + (1-A_{uv})(1-\widehat{P}_{uv}))$

\If{$Likelihood< bestLikelihood$}
 \State$bestLikelihood \gets Likelihood$
 \State $bestParameters \gets \{\widehat{\alpha}, \widehat{\beta}, \widehat{\sigma}, \widehat{\rho}, \widehat{\Psi}\}$
\EndIf
    \end{algorithmic}
    \label{alg:updateP}
\end{algorithm}

\begin{algorithm}
\caption{Update $\Psi_u$ Procedure}
\label{a:update-Psi}
    \begin{algorithmic}
\For{$u$ from 1 to $|\widehat{i}|$}
    \For{$v$ from 1 to $|\widehat{j}|$}
 \State $Estimate_{v} \gets \widehat{\alpha}\left(\Phi_1\left(\frac{1}{\sqrt{1+\widehat{\sigma}^2}}  \Phi_1^{-1}(H_{\widehat{\rho}}(var, \widehat{\Psi}_v))+ \frac{\widehat{\sigma}}{\sqrt{1+\widehat{\sigma}^2}} \che_{u,v}\right)\right) + \widehat{\beta}$ \\

\EndFor
 \State $temp_u \gets \argmin{var}-\left(\sum_{v} A_{uv}\log{Estimate_v} + (1-A_{uv})\log{(1-Estimate_v)}\right)$ \\
 
 \State $Estimate2_{uv} \gets \widehat{\alpha}\left(\Phi_1\left(\frac{1}{\sqrt{1+\widehat{\sigma}^2}}  \Phi_1^{-1}(H_{\widehat{\rho}}(temp_u, \widehat{\Psi}_v))+ \frac{\widehat{\sigma}}{\sqrt{1+\widehat{\sigma}^2}} \che_{u,v}\right)\right) + \widehat{\beta}$
\EndFor

\If{$-(\sum_{u}\sum_{v} A_{uv}\log{Estimate2_{uv}} + (1-A_{uv})\log{(1-Estimate2_{uv})}) < -(\sum_{u}\sum_{v} A_{uv}\log{\widehat{P}_{uv}} + (1-A_{uv})\log{(1-\widehat{P}_{uv})}) $}\\
\State $\{\widehat{\Psi}_u\} \gets \{temp_u\}$\\
\State  $\widehat{P} \gets Estimate2$ 
\EndIf
\end{algorithmic}
\end{algorithm}

\begin{algorithm}
\caption{Update Local Parameters Procedure}
\label{a:updateLocalparams}
    \begin{algorithmic}
     \State $\{Estimate_{uv}\} \gets var1\left(\Phi_1\left(\frac{1}{\sqrt{1+var3^2}}  \Phi_1^{-1}(H_{var4}(\widehat{\Psi}_u, \widehat{\Psi}_v))+ \frac{var3}{\sqrt{1+var3^2}} \che_{u,v}\right)\right) + var2$\\
     
    \State $\{temp_{\alpha}, temp_{\beta},temp_{\sigma}, temp_{\rho}\} \gets \argmin{var1, var2, var3, var4}  -(\sum_{u}\sum_{v} A_{uv}\log{Estimate_{uv}} + (1-A_{uv})\log{(1-Estimate_{uv})})$\\
    
    \State  $\{Estimate2_{uv}\} \gets temp_{\alpha}\left(\Phi_1\left(\frac{1}{\sqrt{1+temp_{\sigma}^2}}  \Phi_1^{-1}(H_{temp_{\rho}}(\widehat{\Psi}_u, \widehat{\Psi}_v))+ \frac{temp_{\sigma}}{\sqrt{1+temp_{\sigma}^2}} \dot{\epsilon}_{uv}\right)\right) + temp_{\beta}$ \\

\If{$-(\sum_{u}\sum_{v} A_{uv}\log{Estimate2_{uv}} + (1-A_{uv})\log{(1-Estimate2_{uv})}) < -(\sum_{u}\sum_{v} A_{uv}\log{\widehat{P}_{uv}} + (1-A_{uv})\log{(1-\widehat{P}_{uv})}) $}\\ 
\State $\{\widehat{\alpha}, \widehat{\beta},\widehat{\sigma}, \widehat{\rho}\} \gets \{temp_{\alpha}, temp_{\beta},temp_{\sigma}, temp_{\rho}\}$ \\
\State $\widehat{P} \gets Estimate2$
 
\EndIf

    \end{algorithmic}
\end{algorithm}

\newpage 
\section{Adjusted Rand Index Tables}

\begin{table}[!ht] \label{t:sim-ARI}
    \centering
    \begin{tabular}
{|p{1.6cm}|p{1.5cm}|p{1.6cm}|p{1.5cm}|p{1.5cm}|p{1.5cm}|p{1.5cm}|p{1.5cm}|p{1.5cm}|p{1.5cm}|}
        \hline
        	&	\textbf{Truth}	&	\textbf{ACRONYM}	&	\textbf{\texttt{nett} 5}	&	\textbf{\texttt{randnet} spectral 5}	&	\textbf{\texttt{randnet} spherical 5}	&	\textbf{\texttt{nett} 16}	&	\textbf{\texttt{randnet} spectral 16}	&	\textbf{\texttt{randnet} spherical 16}	\\ \hline
\textbf{Truth}	&	1.000	&	1.000	&	0.465	&	0.468	&	0.497	&	0.523	&	0.518	&	0.528	\\ \hline
\textbf{ACRONYM}	&	1.000	&	1.000	&	0.465	&	0.468	&	0.497	&	0.523	&	0.518	&	0.528	\\ \hline
\textbf{\texttt{nett} 5}	&	0.465	&	0.465	&	1.000	&	0.951	&	0.732	&	0.369	&	0.361	&	0.382	\\ \hline
\textbf{\texttt{randnet} spectral 5}	&	0.468	&	0.468	&	0.951	&	1.000	&	0.747	&	0.366	&	0.365	&	0.385	\\ \hline
\textbf{\texttt{randnet} spherical 5}	&	0.497	&	0.497	&	0.732	&	0.747	&	1.000	&	0.368	&	0.368	&	0.388	\\ \hline
\textbf{\texttt{nett} 16}	&	0.523	&	0.523	&	0.369	&	0.366	&	0.368	&	1.000	&	0.779	&	0.790	\\ \hline
\textbf{\texttt{randnet} spectral 16}	&	0.518	&	0.518	&	0.361	&	0.365	&	0.368	&	0.779	&	1.000	&	0.813	\\ \hline
\textbf{\texttt{randnet} spherical 16}	&	0.528	&	0.528	&	0.382	&	0.385	&	0.388	&	0.790	&	0.813	&	1.000	\\ \hline

    \end{tabular}\vspace{.05in}
    \caption{Table containing the Adjusted Rand Index Values of the different community detection approaches used on the simulated dataset from Section \ref{s:simulation}.}  
\end{table}

\begin{table}[b!] \label{t:Congress-ARI}
    \centering
    \begin{tabular}
{|p{1.8cm}|p{1.8cm}|p{1.6cm}|p{1.5cm}|p{1.5cm}|p{1.5cm}|}
        \hline

	&	\textbf{ACRONYM}	&	\textbf{\texttt{nett} }	&	\textbf{\texttt{randnet} spectral }	&	\textbf{\texttt{randnet} spherical }	&	\textbf{Leiden}\\ \hline
\textbf{ACRONYM}	&	1.000	&	0.351	&	0.318	&	0.414	&	0.839 \\\hline
\textbf{\texttt{nett} }	&	0.351	&	1.000	&	0.777	&	0.540	&	0.392 \\ \hline
\textbf{\texttt{randnet} spectral }	&	0.318	&	0.777	&	1.000	&	0.452	&	0.353\\ \hline
\textbf{\texttt{randnet} spherical }	&	0.414	&	0.540	&	0.452	&	1.000	&	0.452 \\ \hline
\textbf{Leiden}	&	0.839	&	0.392	&	0.353	&	0.452	&	1.000 \\ \hline
\end{tabular}\vspace{.05in}
    \caption{Table containing the Adjusted Rand Index Values of the different community detection approaches for the Congressional Twitter Dataset}  
\end{table}

\begin{table}[t!] \label{t:Retina-ARI}
    \centering
    \begin{tabular}
{|p{1.8cm}|p{1.8cm}|p{1.6cm}|p{1.5cm}|p{1.5cm}|p{1.5cm}|p{1.8cm}|p{1.8cm}|}
\hline
			
			& \textbf{ACRONYM}	&	\textbf{\texttt{nett} }	&	\textbf{\texttt{randnet} spectral }	&	\textbf{\texttt{randnet} spherical }	&	\textbf{Leiden}	&	\textbf{Coarse Type 1}	&	\textbf{Coarse Type 2}	\\ \hline
\textbf{ACRONYM}		&	1.000	&	0.276	&	0.304	&	0.378	&	0.329	&	0.004	&	0.003	\\ \hline
\textbf{nett}		&	0.276	&	1.000	&	0.673	&	0.414	&	0.155	&	0.001	&	0.002	\\ \hline
\textbf{\texttt{randnet} spectral }		&	0.304	&	0.673	&	1.000	&	0.468	&	0.166	&	0.000	&	0.001	\\ \hline
\textbf{\texttt{randnet} spherical }		&	0.378	&	0.414	&	0.468	&	1.000	&	0.183	&	0.040	&	0.029	\\ \hline
\textbf{Leiden}		&	0.329	&	0.155	&	0.166	&	0.183	&	1.000	&	0.000	&	0.000	\\ \hline
\textbf{Coarse Type 1}		&	0.004	&	0.001	&	0.000	&	0.040	&	0.000	&	1.000	&	1.000
	\\ \hline
\textbf{Coarse Type 2}		&	0.003	&	0.002	&	0.001	&	0.029	&	0.000	&	1.000	&	1.000	\\ \hline

\end{tabular}\vspace{.05in}
    \caption{Table containing the Adjusted Rand Index Values of the different community detection approaches for the Mouse Retina Dataset. The Coarse Type is given in the original dataset, where Coarse Type 1 treats 239 NA entries treated as missing, while Coarse Type 2 treats NA entries as a different subtypes.}  
\end{table}

\end{document}